\documentclass[notitlepage,twocolumn,prl,nobalancelastpage,amssymb,superscriptaddress,showpacs,longbibliography]{revtex4-1}
\usepackage{graphicx}
\usepackage{amsmath}
\usepackage{amssymb}
\usepackage[colorlinks=true,citecolor=blue,linkcolor=blue]{hyperref}
\usepackage{float}
\usepackage{lipsum}
\usepackage{comment}
\usepackage{cancel}
\usepackage[dvipsnames]{xcolor}
\definecolor{IEcolor}{RGB}{255, 153, 51}

\definecolor{CYcolor}{RGB}{50, 100, 50}

\newcommand{\Ch}[1]{\textcolor{black}{#1}} 

\begin{document}

\title{Optical Control of Slow Topological Electrons in Moir{\'e} Systems}

\author{Christopher Yang}
\address{Department of Physics, IQIM, California Institute of Technology, Pasadena, CA 91125, USA}
\author{Iliya Esin}
\address{Department of Physics, IQIM, California Institute of Technology, Pasadena, CA 91125, USA}
\author{Cyprian Lewandowski}
\address{Department of Physics, IQIM, California Institute of Technology, Pasadena, CA 91125, USA}
\address{National High Magnetic Field Laboratory, Tallahassee, Florida, 32310, USA}
\address{Department of Physics, Florida State University, Tallahassee, Florida 32306, USA}
\author{Gil Refael}
\address{Department of Physics, IQIM, California Institute of Technology, Pasadena, CA 91125, USA}

\begin{abstract}
Floquet moir{\'e} materials possess optically-induced flat-electron bands with steady-states sensitive to drive parameters. Within this regime, we show that strong interaction screening and phonon bath coupling can overcome enhanced drive-induced heating. In twisted bilayer graphene (TBG) irradiated by a terahertz-frequency continuous circularly polarized laser, the extremely slow electronic states enable the drive to control the steady state occupation of high-Berry curvature electronic states. In particular, above a critical field amplitude, high-Berry-curvature states exhibit a slow regime where they decouple from acoustic phonons, allowing the drive to control the anomalous Hall response. Our work shows that the laser-induced control of topological and transport physics in Floquet TBG are measurable using experimentally available probes.
\end{abstract}

\maketitle

\maketitle
\begin{figure}[t]
    \centering
    \includegraphics[width=\linewidth]{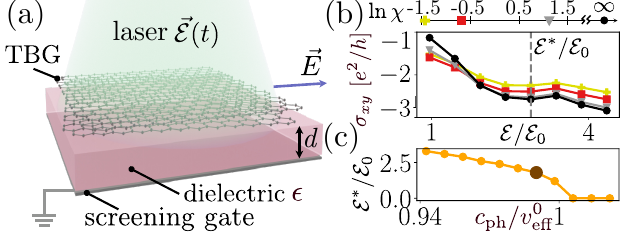}
    \caption{(a) {Schematic experimental design. Circularly polarized laser induces non-trivial Berry-curvature in the \Ch{narrow} bands \Ch{(see Fig. \ref{fig:bandstructure}(b-c))}, resulting in an anomalous Hall conductivity $\sigma_{xy}$.} \Ch{TBG lies on top of a dielectric} and metallic gate \Ch{that screen electron-electron} interactions. (b) Anomalous Hall conductivity vs. drive amplitude $\mathcal{E}$ \Ch{for $\zeta \approx 0.5$ and various values of $\chi$ indicated on the scale (see below Eqs. \ref{eq:fbe} and \ref{eq:phenom-f-kp} for definitions of $\zeta$ and $\chi$). The \Ch{$\sigma_{xy}$}} features a rapid drop with $\mathcal{E}$ below the critical amplitude $\mathcal{E}^*$  (dashed line). Here, \Ch{$\mathcal{E}_0 = \hbar v_F / (e L_M^2) \approx 7.2 \times 10^{4} \ \mathrm{V/m}$.} (c) \Ch{C}ritical amplitude vs. $c_{\text{ph}}/v_{\text{eff}}^0$, where $v_{\text{eff}}^0 = v_{\text{eff}}(0)$ is an effective electron velocity defined in the text. \Ch{E}nlarged red circle: $\mathcal{E}^*$ in (b).}
    \label{fig:intro}
\end{figure}

\textit{Introduction}--- Time-periodic fields can drive materials into exotic non-equilibrium phases \cite{Esin2020FloquetNanowires,gyro,fbe_adv, Castro2022FloquetTheory,Fausti2011Light-inducedCuprate,Flaschner2016ExperimentalBand,Nuske2020FloquetSolids,Fazzini2021NonequilibriumLiquids,floquettopo,Vogl_2023,VoglFloquetEnginneering2021,VoglFloquetEngineeringOfTwistedDouble2020,VoglFloquetEngineeringofTopological2021,LiuTerahertzFieldInducedInsulator,PhysRevB.98.045104}, with unconventional transport and optical characteristics \cite{transport,Kumar2020LinearInsulators,Chono2020Laser-inducedDichalcogenides,Dehghani2021Light-inducedEngineering,Shan2021GiantEngineering,Wong2018PullingField,kitagawa,PhysRevB.105.174301} controllable by external parameters. In laser-driven twisted bilayer graphene (TBG) \cite{Katz2020OpticallyGraphene, Li2020Floquet-engineeredGraphene, Topp2019TopologicalGraphene, VoglFloquetEngineeringOf2020,VoglEffectiveFloquet2020}, a flat-band regime with pronounced electron-electron interaction effects is accessible away from the magic angles \cite{macdonald}. Generating low-temperature Floquet states in such a regime requires cooling processes that compensate for strong drive-induced electron-electron heating. A common cooling solution involves coupling Floquet systems to low-temperature phonon baths \cite{fbe_adv,Dehghani2014DissipativeSystems,Dehghani2015Out-of-equilibriumInsulator}. 

We demonstrate that intrinsic electron-phonon coupling in TBG and Coulomb screening can stabilize low-temperature steady-states in Floquet TBG under terahertz (THz) frequency, circularly polarized laser drives. In this steady-state, the drive amplitude controls the filling of electronic states with large Berry curvature, resulting in a highly tunable anomalous conductivity $\sigma_{xy}$ \cite{transport, McIver2020Light-inducedGraphene, Oka2009PhotovoltaicGraphene, Sato2019MicroscopicGraphene, Sato2019Light-inducedDissipation} (Fig. \ref{fig:intro}(a-b)). The ability to tune the Floquet steady-state results from the unique slow electron regime in TBG where phonons travel faster than---and decouple from---many flat band electronic states \cite{Esin2022GeneratingPhaser, Sharma2021CarrierRegime}.

\begin{figure}
    \centering    \includegraphics[width=0.98\linewidth]{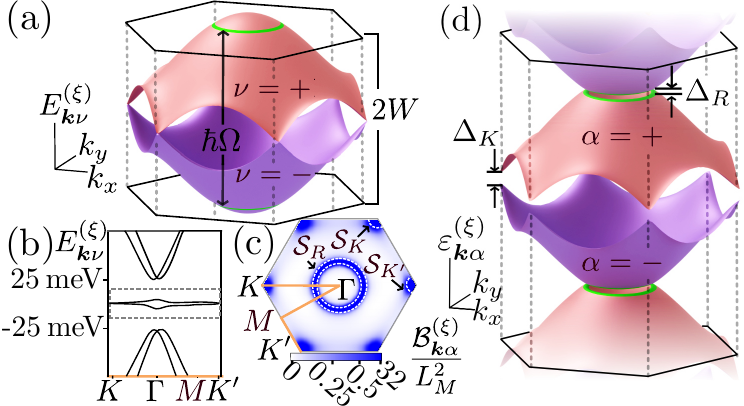}
    \caption{(a) \Ch{Zoom-in on schematic narrow bands} in a moir{\'e} system. Drive with angular frequency $\Omega$ resonantly couples states along resonance rings (green \Ch{curves}). (b) Undriven spectrum of TBG along a line in the Brillouin Zone \Ch{indicated by the orange curve in} (c). Dashed frame encloses optically-active, \Ch{narrow} central bands $\nu = \pm 1$. (c) Berry curvature $\mathcal{B}_{\boldsymbol{k}+}^{(\xi)}$ in the upper Floquet band, with blue color intensity proportional to $\tanh(2\mathcal{B}_{\boldsymbol{k}+}^{(\xi)} / L_M^2)$ (color bar) so $\mathcal{B}_{\boldsymbol{k}+}^{(\xi)}$ peaks are more visible. Dashed lines indicate areas enclosing $\mathcal{B}_{\boldsymbol{k}+}^{(\xi)}$ peaks at the Dirac points and resonance ring. (d) Periodic quasienergy Floquet spectrum of the driven system, having two central bands shown in (a). The Floquet spectrum exhibits the upper (UFB, $\alpha=+$) and lower (LFB, $\alpha=-$) Floquet bands, separated by off-resonant gaps $\Delta_K$ at the Dirac $K$\Ch{,} $K'$ points and a Rabi-like gap $\Delta_R$ along the resonance ring \cite{KarniThroughMaterials, feete-review, Rudner2020BandInsulators}.}
    \label{fig:bandstructure}
\end{figure}

\textit{The system.}---We begin by constructing the time-periodic, interacting Hamiltonian for laser-driven TBG near the charge neutrality point and at a twist angle $\theta$. The single-particle effective Hamiltonian of undriven TBG is $\hat{H}_0 = \sum_{\boldsymbol{k}\nu\xi} E_{\boldsymbol{k}\nu}^{(\xi)} \hat{c}^{(\xi)\dagger}_{\boldsymbol{k}\nu} \hat{c}_{\boldsymbol{k}\nu}^{(\xi)}$, where $\hat{c}_{\boldsymbol{k}\nu}^{(\xi)\dagger}$ creates a Bloch state $|\xi \nu \boldsymbol{k} \rangle$ of crystal momentum $\boldsymbol{k}$, band $\nu$, and energy $E_{\boldsymbol{k}\nu}^{(\xi)}$, \Ch{near} valley index $\xi = \pm 1$ of the single-layer graphene Brillouin zone \cite{macdonald, continuum-koshino}. The index $\nu = \pm$ labels the \Ch{narrow} central particle and hole bands (Fig. \ref{fig:bandstructure}(a, b)) with total bandwidth $W$, \Ch{which are} separated by a large energy gap from all other bands.
We consider a circularly polarized laser of vector potential $\boldsymbol{A}(t) = (\mathcal{E}/\Omega)[\cos(\Omega t) \boldsymbol{\hat{x}} - \sin(\Omega t) \boldsymbol{\hat{y}}]$ with electric field amplitude $\mathcal E$ and angular-frequency $\Omega$, which couples to electrons by minimal coupling $\boldsymbol{k} \to \boldsymbol{k} + e\boldsymbol{A}(t) / \hbar$, resulting in the time-periodic Hamiltonian $\hat{H}_0(t)$. 

The periodic Hamiltonian $\hat{H}_0(t)$ gives rise to Floquet eigenstates $| \Phi_{\boldsymbol{k}\alpha}^{(\xi)} (t) \rangle$ with quasienergies $\varepsilon_{\boldsymbol{k}\alpha}^{(\xi)}$ satisfying  $|\varepsilon_{\boldsymbol{k}\alpha}^{(\xi)}|<\frac{1}{2}\hbar\Omega$. We consider the regime $W \leq \hbar\Omega < 2W$ corresponding to a single photon resonance within the central TBG bands. Specifically, we consider $\Omega = 5 \ \mathrm{meV}/\hbar$ and TBG at a near-magic twist angle of $\theta = 1.13^{\circ}$ whose Fermi velocity $v_F \approx 17 \ \mathrm{km/s}$ (corresponding to $W = 5 \ \mathrm{meV}$ in the Bistritzer-MacDonald model \cite{macdonald,continuum-koshino}) is comparable to phonon speeds in TBG \cite{moirephonon}. The drive mixes the two central bands $\nu = \pm 1$, resulting in quasienergies $\varepsilon_{\boldsymbol{k}\alpha}^{(\xi)}$, with upper and lower Floquet bands denoted by $\alpha = \pm$ (Fig. \ref{fig:bandstructure}(d)) \cite{Katz2020OpticallyGraphene}. The drive opens off-resonant gaps of size $\Delta_K \approx 2e^2 v_F^2 \mathcal{E}^2 /\hbar \Omega^3$ at the Dirac points $K$ and $K'$ of the moir{\'e} Brillouin zone and a Rabi-like gap of $\Delta_R \sim V$ along the resonance ring, which is the ring on the $\boldsymbol{k}$-plane satisfying $E_{\boldsymbol{k}+}^{(\xi)} - E_{\boldsymbol{k}-}^{(\xi)} = \hbar\Omega$ (green rings in Fig. \ref{fig:bandstructure}(a, d)). Here, $v_F$ is the Fermi velocity of the undriven band structure, $V$ is the energy scale of the drive, and the expression for $\Delta_K$ comes from the Van-Vleck perturbative expansion
\cite{Rudner2020BandInsulators, KarniThroughMaterials,Oka2009PhotovoltaicGraphene,Usaj2014IrradiatedInsulator,Aeschlimann2021SurvivalScattering,floquethandbook}. 

The key component for stabilizing Floquet many-body states is the electron coupling to low-temperature longitudinal TBG acoustic phonons:
\begin{equation} \label{eq:h-el-ph}
    \hat{H}_{\text{el-ph}} = \sum_{\substack{\boldsymbol{k},\boldsymbol{q}, \boldsymbol{G} \\ \nu,\nu'.\xi}} {M}_{\boldsymbol{k}\Ch{,\boldsymbol{q}, \boldsymbol{G}}}^{\nu\nu'\xi}  \hat{c}_{\boldsymbol{k}+\boldsymbol{q} + \boldsymbol{G},\nu'}^{(\xi)\dagger} \hat{c}_{\boldsymbol{k}\nu}^{(\xi)}(\hat{b}^{\dagger}_{\boldsymbol{q}} + \hat{b}_{\boldsymbol{-q}}) + \text{h.c.} 
\end{equation}
Here, $\boldsymbol{G}$ is a moir{\'e} Brillouin zone reciprocal lattice vector, and ${M}_{\boldsymbol{k}\Ch{,\boldsymbol{q}, \boldsymbol{G}}}^{\nu\nu'\xi} = {D \sqrt{\hbar c_{\text{ph}} {q}}}/({\sqrt{2A_M\rho}c_{\text{ph}}}) \Ch{\mathcal{W}^{\xi\nu\nu'}_{\boldsymbol{k},\boldsymbol{q}+\boldsymbol{G}}}$ is the matrix element with deformation potential \Ch{$D$}, moir{\'e} unit cell area $A_M = \sqrt{3} L_M^2/2$, lattice vector length $L_M = a / [2 \sin(\theta/2)]$, monolayer graphene density $\rho$, and monolayer lattice vector length $a = 0.246 \ \mathrm{nm}$. The operator $\hat{b}^\dagger_{\boldsymbol{q}}$ creates an acoustic phonon mode of momentum $\boldsymbol{q}$ with amplitude $q$, speed $c_{\text{ph}}$, and energy $\hbar c_{\text{ph}}{q}$. The speed of sound $c_{\text{ph}}$ in TBG is roughly the same as that in monolayer graphene, but the small Brillouin zone in TBG folds the acoustic phonon dispersion into many branches \cite{moirephonon}. The form-factor
$\Ch{\mathcal{W}^{\xi\nu\nu'}_{\boldsymbol{k},\boldsymbol{q}+\boldsymbol{G}} \equiv }\langle \xi \nu'\boldsymbol{k} + \boldsymbol{q} +\boldsymbol{G} |\xi \nu \boldsymbol{k}\rangle$ captures the decreasing coupling of electrons to folded phonon branches with large $\boldsymbol{G}$ \cite{linphonon}. We also include \Ch{electron-electron} interactions:
\begin{equation} \label{eq:h-el-el}
    \Ch{\hat{H}_{\text{el-el}} = \sum_{\substack{\boldsymbol{k}_1,\boldsymbol{k}_2 \\ \boldsymbol{q},\boldsymbol{G} \\ \{\nu_i\},\xi}} {V}_{\boldsymbol{k}_1,\boldsymbol{k}_2,\boldsymbol{q}, \boldsymbol{G}}^{\{\nu_i\}\xi}  \hat{c}_{\boldsymbol{k}_1+\boldsymbol{q},\nu_1}^{(\xi)\dagger} \hat{c}_{\boldsymbol{k}_2 -\boldsymbol{q},\nu_2}^{(\xi)\dagger} \hat{c}_{\boldsymbol{k}_2,\nu_3}^{(\xi)} \hat{c}_{\boldsymbol{k}_1,\nu_4}^{(\xi)},}
\end{equation}
where \Ch{${V}_{\boldsymbol{k}_1,\boldsymbol{k}_2,\boldsymbol{q}, \boldsymbol{G}}^{\{\nu_i\}\xi} = V_{\boldsymbol{q} + \boldsymbol{G}} \mathcal{W}^{\xi\nu_1\nu_4}_{\boldsymbol{k}_1,\boldsymbol{q}+\boldsymbol{G}} \mathcal{W}^{\xi\nu_2\nu_3}_{\boldsymbol{k}_2,-\boldsymbol{q}-\boldsymbol{G}}$, with $i = 1, \dots, 4$, contains the screened Coulomb potential $V_{\boldsymbol{q}} = e^2 / (2 \epsilon_0 q A_M) (1 + \epsilon \coth(qd))^{-1} $ for a gate separated from TBG by a dielectric of permittivity $\epsilon$ and thickness $d$, where $\epsilon_0$ is the vacuum permittivity (Fig. \ref{fig:intro}(a)). }

\begin{figure}
    \centering
    \includegraphics[width=0.9\linewidth]{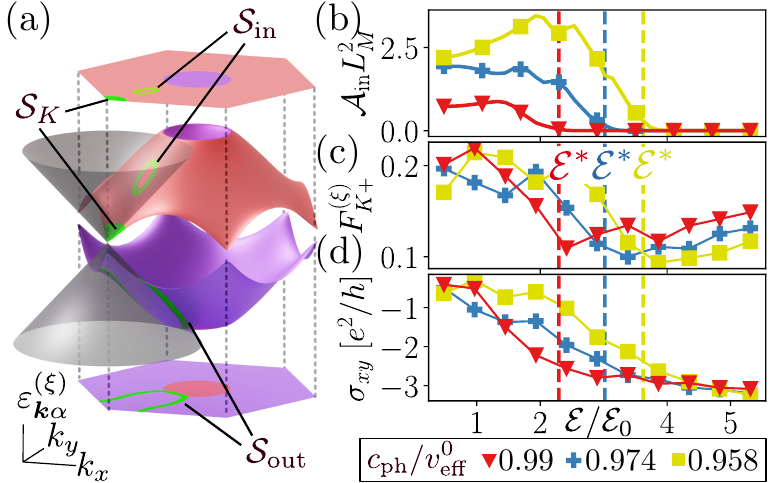}
    \caption{
    (a) Schematics of the Floquet spectrum and one of the phonon light-cones originating from the area $\mathcal{S}_K$ in the UFB. The intersection between the UFB (LFB) and all cones centered in $\mathcal{S}_K$ form $\mathcal{S}_{\text{in}}$ ($\mathcal{S}_{\text{out}}$). As $\mathcal{E} \to \mathcal{E}^*$, the area of $\mathcal{S}_{\text{in}}$ vanishes. (b-d) Numerical verification of the phenomenological model. (b) Area of $\mathcal{S}_{\text{in}}$, $\mathcal{A}_{\text{in}}$, vs. $\mathcal{E}$ for three values of $c_{\text{ph}}/v_{\text{eff}}^0$. (c) Average occupation in $\mathcal{S}_{K}$. (d) Anomalous Hall conductivity $\sigma_{xy}$ for same parameters as (b, c). At $\mathcal{E}^*$ (dashed lines), $\mathcal{A}_{\text{in}}$, $F_{K+}^{(\xi)}$, and $\sigma_{xy}$ \Ch{sharply change}.}
    \label{fig:phenomenology}
\end{figure}

\begin{figure}
    \centering
    \includegraphics[width=\linewidth]{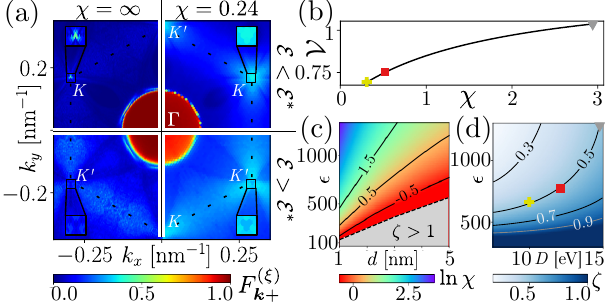}
    \caption{(a) Left column: steady-state occupation of the UFB when $\chi = \infty$ \Ch{(calculated on a $163\times 163$ momentum grid)}. Right column: steady-state occupation \Ch{when $\chi = 0.24$ \Ch{(calculated on a $73\times 73$ momentum grid)}. Bottom row: strong-drive case ($\mathcal{E} = 4.3\mathcal{E}_0 > \mathcal{E}^*$). Top row: weak-drive case ($\mathcal{E} = 0.97\mathcal{E}_0 < \mathcal{E}^*$).} \Ch{Zoom-in boxes: the $K$, $K'$ points have reduced occupation when $\mathcal{E}>\mathcal{E}^*$ relative to when $\mathcal{E} < \mathcal{E}^*$.} \Ch{(b) Visibility $\mathcal{V}$ vs. $\chi$. (c) Value of $\chi$ for various $\epsilon$ and gate distances $d$.  (d) Value of $\eta$ for various $\epsilon$ and deformation potentials $D$, with $d = 4 \ \mathrm{nm}$. Points in (b, d): parameters used in Fig. \ref{fig:intro}(b).} }
    \label{fig:interactions}
\end{figure}

We focus on electron dynamics in its Floquet basis, treating interactions $\hat{H}_{\text{el-ph}}$ and $\hat{H}_{\text{el-el}}$ as weak perturbations scattering electrons between single-particle Floquet states \cite{fbe_adv, fbe_orig, gyro, Sato2019MicroscopicGraphene, Esin2020FloquetNanowires}. The occupation probability $F_{\boldsymbol{k}\alpha}^{(\xi)}(t) = \langle \hat{f}^{(\xi)\dagger}_{\boldsymbol{k}\alpha} (t) \hat{f}_{\boldsymbol{k}\alpha}^{(\xi)} (t) \rangle$ is described by the Floquet-Boltzmann Equation (FBE) \cite{fbe_adv, fbe_orig,PhysRevA.92.062108,PhysRevE.79.051129},
\begin{equation} \label{eq:fbe}
    \dot F_{\boldsymbol{k}\alpha}^{(\xi)}(t) = I^{\text{el-ph}}_{\boldsymbol{k}\alpha}[\{F_{\boldsymbol{k}\alpha}^{(\xi)}(t)\}] + I^{\text{el-el}}_{\boldsymbol{k}\alpha}[\{F_{\boldsymbol{k}\alpha}^{(\xi)}(t)\}].
\end{equation}
Here, $\hat{f}^{(\xi)\dagger}_{\boldsymbol{k}\alpha}(t)$ creates a single-particle electron state $|\Phi_{\boldsymbol{k}\alpha}^{(\xi)}(t)\rangle$, and $I^{\text{el-ph}}_{\boldsymbol{k}\alpha}$ and $I^{\text{el-el}}_{\boldsymbol{k}\alpha}$ are respectively the electron-phonon and electron-electron collision integrals, evaluated by the Fermi golden rule (see Supp. Mat. for FBE details \cite{Yang2022SupplementalMaterials}). The steady-state distribution yields $\dot F_{\boldsymbol{k}\alpha}^{(\xi)}= 0$, and \Ch{$\langle \hat{f}^{(\xi)\dagger}_{\boldsymbol{k}\alpha} (t) \hat{f}_{\boldsymbol{k}\alpha'}^{(\xi)} (t) \rangle$ is supressed for $\alpha \neq \alpha'$ when $1/\tau_{\boldsymbol{k}}^{\text{tot}} \equiv 1/\tau_{\boldsymbol{k}}^{\text{el}} + 1/\tau_{\boldsymbol{k}}^{\text{ph}} \ll \Delta \varepsilon_{\boldsymbol{k}}/\hbar$, where $\tau_{\boldsymbol{k}}^{\text{el}}$ and $ \tau_{\boldsymbol{k}}^{\text{ph}}$ are the interband electron-electron and electron-phonon scattering times, respectively, and $\Delta \varepsilon_{\boldsymbol{k}} = \min_{n\in \mathbb{Z}} |\varepsilon_{\boldsymbol{k}+} - \varepsilon_{\boldsymbol{k}-} + n\hbar\Omega|$ \cite{fbe_orig, Kohn2001PeriodicThermodynamics,PhysRevE.79.051129}. Because $\Delta \varepsilon_{K} =2 \Delta_K$ is minimal, the condition is equivalently $\zeta \equiv \hbar/(2\Delta_K \tau^{\text{tot}}_{K}) \ll 1$ (see Fig. \ref{fig:interactions}(d)). In Figs. \ref{fig:intro}(b) and \ref{fig:interactions}(c-d), we show the maximal $\zeta$ across fields $\mathcal{E}$ plotted in Figs. \ref{fig:intro} and \ref{fig:phenomenology}.}

\textit{Transport properties.}---To probe the electronic dynamics induced by the laser, we study the anomalous conductivity in the steady-state of the system \cite{McIver2020Light-inducedGraphene, Oka2009PhotovoltaicGraphene,transport,Sato2019MicroscopicGraphene,Sato2019Light-inducedDissipation,Dehghani2015Out-of-equilibriumInsulator,Chandran2016Interaction-stabilizedModel,Tomadin2011NonequilibriumSystem,Dehghani2016FloquetProduction}
\begin{equation} \label{eq:conductivity}
     \sigma_{xy} = \frac{2e^2}{\Ch{\hbar}}  \sum_{\alpha, \xi = \pm} \int {d^2\boldsymbol{k}} \mathcal{B}_{\boldsymbol{k}\alpha}^{(\xi)} F_{\boldsymbol{k}\alpha}^{(\xi)},
\end{equation} 
which averages the product of Berry curvature \cite{Rudner2020BandInsulators, transport, berry}
\begin{equation}
    \mathcal{B}_{\boldsymbol{k}\alpha}^{(\xi)} = \frac{\Omega}{\pi} \int_0^{2\pi/\Omega} dt \  \text{Im}\langle \partial_{k_x}\Phi_{\boldsymbol{k}\alpha}^{(\xi)} (t) | \partial_{k_y} \Phi_{\boldsymbol{k}\alpha}^{(\xi)} (t) \rangle, 
\end{equation}
and the steady-state fillings, $F_{\boldsymbol{k}\alpha}^{(\xi)}$. Without the drive, TBG has fragile topology with $\sigma_{xy}=0$ at charge neutrality \cite{Nuckolls2020StronglyGraphene,Repellin2020ChernTransition,Xie2021FractionalGraphene}. The circularly polarized laser breaks time-reversal symmetry between the valleys $\xi = \pm 1$, opens Haldane gaps in each valley, and produces nonzero $\sigma_{xy}$. 

Our main finding is that $\sigma_{xy}$ can be controlled by the field strength. It features a rapid drop as a function of the amplitude of the drive, $\mathcal{E}$, near the critical amplitude $\mathcal{E}^*$ (Fig. \ref{fig:intro}(b)). This strong dependence on the external field indicates profound changes in the electronic steady-state distribution as the drive amplitude changes across $\mathcal{E} = \mathcal{E}^*$. Furthermore, this strong amplitude-dependence arises only when the undriven effective electronic velocity $v_{\text{eff}}^0$ is close to $c_{\text{ph}}$ in TBG (Fig. \ref{fig:intro}(c)), a condition unique to TBG near the ``slow-electron'' regime \cite{Esin2022GeneratingPhaser, Sharma2021CarrierRegime}. 

\textit{Phenomenological analysis.}---We explain the origin of the strong dependence of $\sigma_{xy}$ on the drive amplitude near $\mathcal{E} = \mathcal{E}^*$ (Fig. \ref{fig:intro}(b)) by focusing on key processes affecting $\sigma_{xy}$, which involve momentum states (the $K$ and $K'$ points and resonance ring, see Fig. \ref{fig:bandstructure}(c)) with large Berry curvature $\mathcal{B}_{\boldsymbol{k}\alpha}^{(\xi)}$. We assume that the steady-state occupation of the upper Floquet band (UFB, $\alpha = +$) and valley index $\xi$ near $K$ are uniform, $F_{\boldsymbol{k}+}^{(\xi)} = F_{K+}^{(\xi)}$, for $\boldsymbol{k} \in \mathcal{S}_{K}$, where $\mathcal{S}_{K}$ is a small circle enclosing the full-width half maximum of the Berry curvature peak at $K$ (Fig. \ref{fig:bandstructure}(c)). 

The steady-state occupation emerges as a balance between the total incoming \Ch{rate $\dot{F}_{K+}^{(\xi)}|_{\text{in}}$ into $\mathcal{S}_{K}$} and outgoing rate $\dot{F}_{K+}^{(\xi)}|_{\text{out}}$ \Ch{from $\mathcal{S}_{K}$}. Single phonon emission connecting the UFB $\mathcal{S}_{\text{in}}$ \Ch{(see Fig. \ref{fig:phenomenology}(a))} with $\mathcal{S}_{K}$ is the dominant contribution to  $\dot{F}_{K+}^{(\xi)}|_{\text{in}}$. The two regions are connected by the phonon light-cone (\Ch{see} Fig \ref{fig:phenomenology}(a)). This rate is $\dot F_{K+}^{(\xi)}|_{\text{ph,in}} \approx\mathcal{R}_{\text{in}}(1-F_{K+}^{(\xi)})F_{\text{in}}^{(\xi)}$, where $F_{\text{in}}^{(\xi)}$ is the average UFB occupation in $\mathcal{S}_{\text{in}}$, and $\mathcal{R}_{\text{in}}$ is the average intrinsic scattering rate. Importantly, $\mathcal{R}_{\text{in}}$ is proportional to the momentum-space area of $\mathcal{S}_{\text{in}}$, denoted $\mathcal{A}_{\text{in}}$, estimated by counting  the UFB states that may scatter to $\mathcal{S}_{K}$ by electron-phonon interactions. Hence, $\mathcal{S}_{\text{in}}$ is the intersection between the UFB and phonon light-cones originating anywhere within $\mathcal{S}_{K}$ (Fig. \ref{fig:phenomenology}(a)). As $\Delta_R$ and $\Delta_K$ widen with $\mathcal{E}$, the Floquet bands become narrower \cite{Dehghani2021Light-inducedEngineering,Li2020Floquet-engineeredGraphene,Katz2020OpticallyGraphene}, and $\mathcal{A}_{\text{in}}$ shrinks, vanishing at $\mathcal{E} = \mathcal{E}^*$ (Fig. \ref{fig:phenomenology}(b)). The critical strength $\mathcal{E}^*$ is defined by $v_{\text{eff}}(\mathcal{E}^*) = c_{\text{ph}}$, where $v_{\text{eff}}(\mathcal{E}) = \max_{\boldsymbol{k}'} ({\varepsilon_{\boldsymbol{k}'+}^{(\xi)} - \varepsilon_{\boldsymbol{K}+}^{(\xi)}})/{|\boldsymbol{k}'-\boldsymbol{K}|}$ is the electronic velocity near the $K$ point. By estimating $v_{\text{eff}}(\mathcal{E})$, one finds that $\mathcal{E}^* \propto [1-c_{\text{ph}} / v_{\text{eff}}^0]^{\gamma}$ for small $1-c_{\text{ph}} / v_{\text{eff}}^0$, where $\gamma$ depends on the quasienergy structure and $v_{\text{eff}}^0 \equiv v_{\text{eff}}(0)$. One can also show $\mathcal{A}_{\text{in}}\propto \max(\mathcal{E}-\mathcal{E}^*, 0)$ as $\mathcal{E} \to \mathcal{E}^*$. (See Supp. Mat. \cite{Yang2022SupplementalMaterials}.)

Similarly, the phonon-mediated outgoing rate is $\dot F_{K+}^{(\xi)} |_{\text{ph,out}} \approx\mathcal{R}_{\text{out}} F_{K+}^{(\xi)} (1-F_{\text{out}}^{(\xi)})$, where $F_{\text{out}}^{(\xi)}$ is the lower Floquet band (LFB, $\alpha = -$) average occupation in $\mathcal{S}_{\text{out}}$, and $\mathcal{R}_{\text{out}}$ is the average intrinsic rate, proportional to $\mathcal{A}_{\text{out}} = \int_{\mathcal{S}_{\text{out}}} d^2 \boldsymbol{k}$, where $\mathcal{S}_{\text{out}}$ is the momentum region enclosing intersections between the LFB with phonon light cones originating from states in $\mathcal{S}_{K}$ (\Ch{see} Fig. \ref{fig:phenomenology}(a)). However, unlike $\mathcal{A}_{\text{in}}$, $\mathcal{A}_{\text{out}}$ does not vanish as $\mathcal{E} \to \mathcal{E}^*$ and instead expands as $\mathcal{E}$ increases. 

\Ch{Electron-electron} interactions and photon-mediated Floquet-Umklapp (FU) processes introduce additional terms in the rate equation depending smoothly on $\mathcal{E}$ and roughly uniformly-spread in momentum. We thus \Ch{include} an incoming $\dot F_{K+}^{(\xi)}|_{\text{r,in}} = \Gamma_{\text{in}}(1-F_{K+}^{(\xi)})$ and outgoing rate $\dot F_{K+}^{(\xi)}|_{\text{r,out}} = \Gamma_{\text{out}}F_{K+}^{(\xi)}$ with \Ch{$\Gamma_{\text{in}/\text{out}} \equiv \Gamma_{\text{in/out}}^{\text{ph}}+\Gamma_{\text{in/out}}^{\text{el}}$, where $\Gamma_{\text{in/out}}^{\text{el(ph)}}$, are rates of electron-electron (electron-phonon FU) processes.} The strength of FU processes is weaker than $\mathcal{R}_{\text{out}}$ by factors of $\Ch{\approx}(v_F e \mathcal{E}/\Omega^2)^{2n}$, where $|n|>1$ is the number of photons emitted or absorbed \cite{fbe_adv}. \Ch{FU} processes also impart large phonon momentum transfers \Ch{that} the form-factor in Eq. \ref{eq:h-el-ph} suppresses. 

\Ch{In the steady-state, $\dot{F}_{K+}^{(\xi)}|_{\text{in}} =\dot F_{K+}^{(\xi)}|_{\text{ph,in}} + \dot F_{K+}^{(\xi)}|_{\text{r,in}}$ and $\dot{F}_{K+}^{(\xi)}|_{\text{out}} = \dot F_{K+}^{(\xi)}|_{\text{ph,out}} + \dot F_{K+}^{(\xi)}|_{\text{r,out}}$ are equal, and}
\begin{equation} \label{eq:phenom-f-kp}
    F_{K+}^{(\xi)} = \frac{\mathcal{R}_{\text{in}} F_{\text{in}}^{(\xi)} + \Gamma_{\text{in}}}{ \mathcal{R}_{\text{in}} F_{\text{in}}^{(\xi)} + \mathcal{R}_{\text{out}} \Ch{(1-F_{\text{out}}^{(\xi)})} + \Gamma_{\text{in}} + \Gamma_{\text{out}}}.
\end{equation}
\Ch{Note that $F_{\text{in}}^{(\xi)}, 1-F_{\text{out}}^{(\xi)} \neq 0$ due to electron (hole) excitations in the UFB (LFB) generated by FU processes.} Since $\mathcal{R}_{\text{in}}\propto\mathcal{A}_{\text{in}}$, \Ch{$\mathcal{R}_{\text{in}}$} decreases as a function of $\mathcal{E}$\Ch{, shrinking to zero} for $\mathcal{E} \geq \mathcal{E}^*$ (\Ch{see} Fig. \ref{fig:phenomenology}(b) for numerical verification). We expect a similar $\mathcal{E}$-dependence of \Ch{$F_{K+}^{(\xi)}$} and $\sigma_{xy}$, yet smeared by additional scattering rates appearing in Eq. \ref{eq:phenom-f-kp}, as \Ch{verified} numerically in Fig. \ref{fig:intro}(b). \Ch{Additionally, Eq. \ref{eq:phenom-f-kp} elucidates the dependence of $F^{(\xi)}_{K+}$ on the ratio $\chi \equiv \tau^{\text{el}}_K/\tau^{\text{ph}}_K \approx \mathcal{R}_{\text{out}}/\Gamma_{\text{out}}^{\text{el}} \approx \mathcal{R}_{\text{out}}/\Gamma_{\text{in}}^{\text{el}}$ (see Fig. \ref{fig:intro}(b)), with $F^{(\xi)}_{K+} \to 0.5$ as $\chi \to 0$. In Figs. \ref{fig:intro}(b) and \ref{fig:interactions}(b-c), we display $\chi$ evaluated at the amplitude $\mathcal{E}$ where $\zeta$ is fixed.}

\Ch{A similar rate equation can be derived for the occupation probability of holes in the LFB. Due to the emergent, approximate anti-unitary particle-hole symmetry \cite{PhysRevB.103.205413, PhysRevB.103.205412, PhysRevLett.123.036401} at charge neutrality that is preserved by the drive, the transition rates in the LFB are roughly similar to those in the UFB, leading to approximately equal electron and hole occupations near the Dirac points in the UFB and LFB ($F_{K+}^{(\xi)} \approx 1-F_{K'-}^{(\xi)}$). Notice that the signs of the Berry curvatures near the Dirac points in the LFB and UFB are opposite, resulting in constructive contributions of electron and hole populations to $\sigma_{xy}$.} \Ch{Thus, we can reproduce qualitatively the \Ch{sharp change} of $\sigma_{xy}$ with $\mathcal{E}$ in Fig. \ref{fig:phenomenology} \cite{Katz2020OpticallyGraphene}}. \Ch{Occupations in the} resonance ring vicinity (Fig. \ref{fig:bandstructure}(c)) yield a similar $\mathcal{E}$-dependence, but with a much lower critical field \Ch{(not visible for $\mathcal{E}$ plotted in Figs. \ref{fig:intro} and \ref{fig:phenomenology})} due to different effective electronic velocities near the resonance ring. 

\textit{Numerical analysis.}---The results in Fig. \ref{fig:phenomenology}(b-d) utilized a simplified toy model describing TBG as a tight-binding hexagonal lattice, similar to graphene \cite{floquettbg}, but with parameters tuned to match $v_F$ and the Brillouin zone size of TBG. This model misses some subtle details but captures the interplay between electron and phonon velocities and the large Berry curvature at the Dirac points and resonance ring. The model represents only the central $\nu = \pm 1$ bands of the undriven bandstructure, but since the low drive angular frequency $\Omega$ is only resonant to these \Ch{narrow} bands, we can ignore the $|\nu| > 1$ bands---valid when $\theta$ is near the magic angle where the $|\nu| > 1$ and $\nu = \pm 1$ bands are well-separated. In the Supp. Mat. \cite{Yang2022SupplementalMaterials}, we present the numerical analysis of a continuum model without \Ch{electron-electron} interactions \cite{continuum-koshino, macdonald,mp-grid}, which yields qualitatively similar results, \Ch{demonstrating that the controllable $\sigma_{xy}$ is insensitive to model details.} In the toy model, $v_{\text{eff}}^0 = 18.9 \ \mathrm{km/s}$, and we \Ch{vary} $c_{\text{ph}} \in [17.9 \ \mathrm{km/s}, 19.4 \ \mathrm{km/s}]$ \Ch{in Fig. \ref{fig:intro}(c)}. In the range $c_{\text{ph}} < v_{\text{eff}}^0$, the drive induces the regime $c_{\text{ph}} > v_{\text{eff}}(\mathcal{E})$ for $\mathcal{E} > \mathcal{E}^*$. \Ch{To capture the decaying overlap of the wavefunctions for momentum Umklapp transitions, in the toy model, we take $\mathcal{W}^{\xi\nu\nu'}_{\boldsymbol{k},\boldsymbol{q}} \to \langle \xi \nu' \boldsymbol{k} + \boldsymbol{q} |\xi \nu \boldsymbol{k} \rangle e^{-l_w^2 q^2/4}$, with $l_w \approx L_M/(1.5\sqrt{3})$} representing the \Ch{radius} of Wannier orbitals localized to TBG layer alignment sites  \cite{linphonon, Yang2022SupplementalMaterials}. 

First, we show how solving the FBE (Eq. \ref{eq:fbe}) for the steady-state distribution verifies the phenomenological model. Consider the non-interacting limit by solving Eq. \ref{eq:fbe} for $F_{\boldsymbol{k}\alpha}^{(\xi)}$ with $\Ch{\chi} \to \infty$ ($I^{\text{el-el}}_{\boldsymbol{k}\alpha} = 0$). The \Ch{left-half column} of Fig. \ref{fig:interactions}(a) shows the non-interacting steady-state distributions for a phonon bath temperature of $1 \ \mathrm{K}$ and \Ch{$c_{\text{ph}} = 0.99 v_{\text{eff}}^0$} in the $\mathcal{E} > \mathcal{E}^*$ and $\mathcal{E} < \mathcal{E}^*$ cases. \Ch{When} $\mathcal{E} > \mathcal{E}^*$ (left \Ch{bottom} quadrant), the \Ch{Dirac} points have reduced occupations \Ch{(see zoom-in boxes)} relative to \Ch{when} $\mathcal{E} < \mathcal{E}^*$ (left \Ch{top} quadrant), \Ch{because} incoming scattering rates into $\mathcal{S}_{K,K'}$ \Ch{are suppressed} (verifying the phenomenological model). Fig. \ref{fig:phenomenology}(c) shows the occupation near the $K$ point, $F_{K+}^{(\xi)}$, as a function of $\mathcal{E}$ for three values of $c_{\text{ph}} / v_{\text{eff}}^0$ and verifies $\mathcal{A}_{\text{in}}$, $F_{K+}^{(\xi)}$, and $\sigma_{xy}$ \Ch{sharply change} at the same critical \Ch{field} $\mathcal{E} = \mathcal{E}^*$. \Ch{Heating induced by FU processes causes $F_{K+}^{(\xi)}$ to slowly increase with $\mathcal{E} > \mathcal{E}^*$ (see Eq. \ref{eq:phenom-f-kp}).}

Next, we quantify the strength of Coulomb screening necessary to stabilize the steady-state, \Ch{which depends on the balance between electron-phonon cooling processes and electron-electron heating processes.} We include $I^{\text{el-el}}_{\boldsymbol{k}\alpha} \neq 0$ \Ch{by taking finite $\chi$}. On the right\Ch{-half column} of Fig. \ref{fig:interactions}(a), we show the resulting steady-state occupations, which \Ch{is slightly closer to the hot steady-state $F_{\boldsymbol{k}\pm}^{(\xi)}=0.5$ and has more smeared occupations} than the non-interacting case \Ch{(left half of Fig. \ref{fig:interactions}(a)). To quantify the effect of interactions on $\sigma_{xy}$, note that, in Fig. \ref{fig:intro}(b), $\sigma_{xy}$ drops less rapidly with $\mathcal{E} < \mathcal{E}^*$ as $\chi$ decreases. We capture this behavior with the visibility parameter $\mathcal{V} \equiv - {\Ch{\max_{\mathcal{E} < \mathcal{E}^*}}{|\partial_{\mathcal{E}} \sigma_{xy}}| }/[{(e^2/h) / \mathcal{E}_0}]$}. \Ch{Fig. \ref{fig:interactions}(b) demonstrates how $\mathcal{V}$ increases with $\chi$. Lastly, we relate $\chi$ and $\zeta$ to physical parameters in TBG. Fig. \ref{fig:interactions}(c) shows the necessary gate distances $d$ and dielectrics $\epsilon$ to experimentally achieve various values of $\chi$, and Fig. \ref{fig:interactions}(d) shows the values of $\epsilon$ and deformation potentials $D$ satisfying $\zeta < 1$ for $d = 4 \ \mathrm{nm}$. One suitable dielectric is $\mathrm{SrTiO_3}$ with $\epsilon \sim 1600$ at $\Omega = 5 \ \mathrm{meV/\hbar}$ angular frequencies \cite{PhysRevB.94.224515, PhysRevB.24.3086,doi:10.1063/1.363513}; note that surface optical phonons in $\mathrm{SrTiO_3}$ are of higher frequencies than $\Omega$ and would not interact with electrons in TBG via direct (non-FU) scattering processes \cite{so-phonons-srtio3}.}

\textit{Conclusion}---TBG is a remarkable system \Ch{whose} Fermi velocity is comparable to the speed of sound. Upon THz-laser driving, the electronic population dynamics exhibits bottlenecks for electron-phonon scattering into high-Berry curvature Floquet states, strongly affecting the anomalous Hall transport. These bottlenecks can be sensitively controlled by the drive amplitude. If the undriven effective electron speed is faster than sound $v_{\text{eff}}^0>c_{\text{ph}}$, a drive with $\mathcal{E}>\mathcal{E}^*$ induces the opposite regime $v_{\text{eff}}(\mathcal{E})<c_{\text{ph}}$, \Ch{weakening the electron-phonon coupling} \Ch{and} suppress\Ch{ing} the Hall conductivity (Fig. \ref{fig:intro}(b)). We also find that a strong \Ch{$\mathcal{E}$}-dependence of $\sigma_{xy}$ arises for efficient Coulomb screening by a close-by gate or a strong dielectric \cite{coissard_imaging_2022,veyrat_helical,PhysRevLett.26.851}. Experimental advances in Floquet engineering \cite{McIver2020Light-inducedGraphene}, and THz laser sources \cite{Lewis2014ASources,thz-laser}, show that our predicitions should be accessible experimentally.  

Analysis of UV-visible or X-ray driven TBG is a subject of future work, which must account for \Ch{optically-active} dispersive bands \cite{Katz2020OpticallyGraphene,VoglEffectiveFloquet2020}. High-frequency drives could reduce heating, facilitating fewer electron-electron FU processes \cite{fbe_adv} while \Ch{activating} electron-phonon Umklapp cooling processes arising from tightly-localized  Wannier orbitals in TBG \cite{linphonon}. (\Ch{In this work, these cooling processes are} suppressed \Ch{FU} processes.) Another interesting direction involves \Ch{developing a Hartree-Fock treatment for }symmetry-broken phases in the steady-state of strongly coupled TBG \cite{gyro}. We leave these exciting directions to future studies.

\begin{acknowledgments}
We thank Netanel Lindner, Mark Rudner, Or Katz, Gaurav Gupta, Seamus O'Hara, Jason Alicea, Alex Thomson, Felix von Oppen, Kry\v{ s}tof Kol{\'a}\v{r}, \Ch{{\'E}tienne Lantagne-Hurtubise, and Valerio Peri} for valuable discussions. C.Y. gratefully acknowledges support from the DOE NNSA Stewardship Science Graduate Fellowship program, which is provided under cooperative agreement number DE-NA0003960. C.L. acknowledges support by the Gordon and Betty Moore
Foundation’s EPiQS Initiative, Grant GBMF8682, start-up funds from Florida State University and the National High Magnetic Field Laboratory. The National High Magnetic Field Laboratory is supported by the National Science Foundation through NSF/DMR-1644779 and the state of Florida.
G.R. and I.E. are grateful for support from the Simons Foundation and the Institute of Quantum Information and Matter, as well as support from the NSF DMR grant number 1839271. 
This work is supported by ARO MURI Grant No. W911NF-16-1-0361, and was performed in part at Aspen Center for Physics, which is supported by National Science Foundation grant PHY-1607611.
\end{acknowledgments}

%

\end{document}


\widetext
\onecolumngrid

\begin{center}
\textbf{\large Supplemental Material:
\\Optical Control of Slow Topological Electrons in Moir{\'e} Systems}
\\[0.4ex] Christopher Yang, Iliya Esin, Cyprian Lewandowski, and Gil Refael
\end{center}
\par
\setcounter{page}{1}
\twocolumngrid

\setcounter{equation}{0}
\setcounter{figure}{0}
\setcounter{table}{0}
\setcounter{page}{1}
\makeatletter
\renewcommand{\theequation}{S\arabic{equation}}
\renewcommand{\thefigure}{S\arabic{figure}}
\setcounter{secnumdepth}{4}

\section{Details of the Models} \label{sec:models_supp}
In both the toy and continuum models, we take the undriven Hamiltonians $H(\boldsymbol{k})$ and obtain the time-dependent Hamiltonian $H(\boldsymbol{k},t)$ via minimal coupling $\boldsymbol{k} \to \boldsymbol{k} + e \boldsymbol{A}(t) / \hbar$. Here,
\begin{equation} \label{eq:vecpotential}
    \boldsymbol{A}(t) = A [\cos(\Omega t) \boldsymbol{\hat{x}} - \sin(\Omega t) \boldsymbol{\hat{y}}]
\end{equation}
is the magnetic vector potential of \Ch{the} circularly polarized laser. We can expand the time-dependent eigenstates of the Hamiltonian in a Floquet-Bloch basis \cite{floquethandbook}:
\begin{equation}
    | \psi_{\boldsymbol{k}\alpha} (t) \rangle =  e^{-i\varepsilon_{\alpha}^{(\xi)} t/\hbar} |\Phi_{\boldsymbol{k}\alpha}^m (t) \rangle,
\end{equation}
where $|\Phi_{\boldsymbol{k}\alpha}^m (t) \rangle$ is periodic in time ($|\Phi_{\boldsymbol{k}\alpha}^m (t) \rangle = |\Phi_{\boldsymbol{k}\alpha}^m (t+2\pi /\Omega) \rangle$), $\varepsilon_{\alpha}^{(\xi)}$ are the quasienergies plotted in Fig. 2(d), and $\alpha$ enumerates the Floquet quasienergy bands. To determine the Floquet-Bloch basis, it is easiest to expand the time-dependent $|\Phi_{\boldsymbol{k}\alpha}^m (t) \rangle$ in terms of time-independent Fourier harmonics $|\phi^m_{\boldsymbol{k}\alpha}\rangle$,
\begin{equation}
    |\Phi_{\boldsymbol{k}\alpha}^m (t) \rangle = \sum_m e^{-im\Omega t} |\phi^m_{\boldsymbol{k}\alpha}\rangle,
\end{equation}
take a Fourier transform the Hamiltonian,
\begin{equation}
    H(\boldsymbol{k},t) = \sum_m e^{-im\Omega t} H^{(m)}(\boldsymbol{k}),
\end{equation}
and solve the Schrödinger equation in the basis of Floquet harmonics:
\begin{equation} \label{eq:fourier-hamiltonian}
    (\varepsilon_{\alpha}^{(\xi)} + m\hbar\Omega) |\phi^m_{\boldsymbol{k}\alpha}\rangle = \sum_{m'} H^{(m-m')}(\boldsymbol{k}) |\phi^{m'}_{\boldsymbol{k}\alpha}\rangle.
\end{equation}
In the following subsections, we detail the exact form of the Floquet Hamiltonians.

\subsection{Tight binding Floquet toy Hamiltonian} \label{sec:toy-model-details}
We use a rescaled, two-band tight binding model for graphene to replicate the flat conduction and valence bands of TBG. In the rescaled Hamiltonian
\begin{equation}
    H_{\text{toy}}({\boldsymbol{k}}) = \begin{pmatrix}
        0 & h_{\boldsymbol{k}} \\
        h_{\boldsymbol{k}}^* & 0
    \end{pmatrix},
\end{equation}
\begin{equation}
    h_{\boldsymbol{k}} = \frac{W}{3} \sum_{j} e^{i\boldsymbol{k}\cdot\boldsymbol{\delta}_j},
\end{equation}
we choose long hopping vectors 
\begin{equation}
    \boldsymbol{\delta}_j = L_M /\sqrt{3}[\sin(2\pi m/3) \boldsymbol{\hat{x}} + \cos(2\pi m/3) \boldsymbol{\hat{y}}],
\end{equation}
with $L_M = 0.246 \ \mathrm{nm} / (2 \sin \theta/2)$, and a narrow bandwidth $W$. The corresponding rescaled \Ch{energies} and Bloch states are
\begin{equation}
    E_{\nu}(\boldsymbol{k}) = {\nu} |h_{\boldsymbol{k}}|,
\end{equation}
and
\begin{equation}
    |{\nu} \boldsymbol{k} \rangle = \frac{1}{\sqrt{2}} \begin{pmatrix}
        \nu e^{i \mathrm{arg}(h_{\boldsymbol{k}})} \\ 1 
    \end{pmatrix},
\end{equation}
respectively, with ${\nu} = \pm 1$ enumerating the \Ch{narrow} Bloch bands.

Following \Ch{Ref.} \cite{floquettbg}, we perform minimal coupling, which turns the functions $h_{\boldsymbol{k}}$ into time-dependent quantities with Fourier transforms
\begin{align}
\begin{split}
    {h}_{\boldsymbol{k}}^{(n)} &= \frac{1}{2\pi /\Omega} \int_0^{2\pi/\Omega} h_{\boldsymbol{k} + e\boldsymbol{A}(t)/\hbar} e^{-in\Omega t} dt \\
    &= \sum_{j} t e^{i\boldsymbol{k}\cdot\boldsymbol{\delta}_j} e^{in\phi_j} J_n(-\tilde{\mathcal{E}}),
\end{split}
\end{align}
where $\tilde{\mathcal{E}}$ is the dimensionless drive strength 
\begin{equation} \label{eq:dimensionless-drive-strength}
    \tilde{\mathcal{E}} = \frac{eL_M}{\sqrt{3} \hbar} A = \frac{eL_M}{\sqrt{3} \hbar} \frac{\mathcal{E}}{\Omega};
\end{equation}
the phase angles are $\phi_0 = \pi / 2$, $\phi_1 = -5\pi/6$, and $\phi_2 = -\pi/6$; and
\begin{equation}
    J_n(z) = \frac{1}{2\pi i^n} \int_0^{2\pi} e^{iz\cos\theta} e^{in\theta} d\theta.
\end{equation}
The Fourier-transformed Hamiltonian is
\begin{equation}
    H_{\text{toy}}^{(n)}(\boldsymbol{k}) = \begin{pmatrix}
        0 & {h}_{\boldsymbol{k}}^{(n)} \\
        {h}_{\boldsymbol{k}}^{*(n)} & 0
    \end{pmatrix}.
\end{equation}
Note that
\begin{equation}
    {h}_{\boldsymbol{k}}^{*(n)} = \sum_{j} t e^{-i\boldsymbol{k}\cdot\boldsymbol{\delta}_j} e^{in\phi_j} J_n(\tilde{\mathcal{E}})
\end{equation}
is the Fourier transform of the conjugate of $h_{\boldsymbol{k}}$. In simulations, we generally truncate the Fourier Hamiltonian (Eq. \ref{eq:fourier-hamiltonian}) to $-12 \leq m \leq 12$, so that we account for a sufficient number of high-order Floquet-Umklapp processes in the Floquet-Boltzmann equation. \Ch{We do not perform the gauge transformation $h_{\boldsymbol{k}}^{(n)} \to i e^{-i\boldsymbol{k} \cdot \boldsymbol{\delta}_0} h_{\boldsymbol{k}}^{(n)}$ so as to preserve the $C_3$ symmetry of the matrix element in the Floquet-Boltzmann equation (see Eq. \ref{eq:el-ph-m}).}
\begin{figure}
    \centering
    \includegraphics[width=\linewidth]{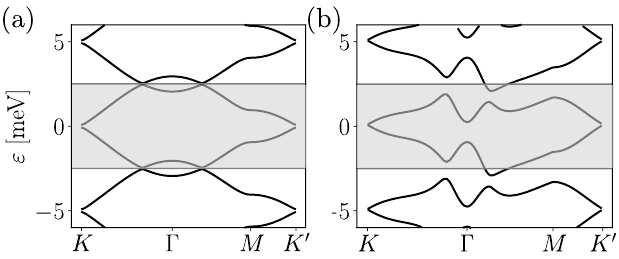}
    \caption{(a) The quasienergy band structure of the toy model with for the parameters used in the main text. (b) The quasienergy band structure of the continuum model at valley $\xi = +1$. In both panels, the first Floquet Brillouin zone is shaded. See Sec. \ref{sec:choice} for details and justification for the parameters we have used.}
    \label{fig:fig_bands}
\end{figure}

\subsection{Continuum Model Floquet Hamiltonian} \label{sec:cont-model-hamiltonian}
The undriven continuum model for TBG \cite{continuum-koshino} describes the bandstructure of TBG near the valley $\xi = \pm 1$ of the monolayer graphene Brillouin zone. Its Hamiltonian
\begin{equation} \label{eq:continuum-hamiltonian}
    H_{\xi}\Ch{(\boldsymbol{k})} = \begin{pmatrix}
        H_1^{\Ch{\xi}}\Ch{(\boldsymbol{k})} & U^\dagger_{\Ch{\xi}} \\
        U_{\Ch{\xi}} & H_2^{\Ch{\xi}}(\Ch{\boldsymbol{k}})
    \end{pmatrix}
\end{equation}
is diagonalized in the basis $\psi_{n\boldsymbol{k}} =(\psi^{A_1}_{n\boldsymbol{k}}, \psi^{B_1}_{n\boldsymbol{k}}, \psi^{A_2}_{n\boldsymbol{k}}, \psi^{B_2}_{n\boldsymbol{k}})^T$ with
\begin{equation}
    \psi^{X}_{n\boldsymbol{k}}(\boldsymbol{r}) = e^{i\boldsymbol{k}_{\Ch{\text{mic}}}\cdot\boldsymbol{r}} \sum_{\boldsymbol{G}} C^X_{n\boldsymbol{k}} (\boldsymbol{G}) e^{i\boldsymbol{G}\cdot\boldsymbol{r}}
\end{equation}
where $X = A_l, B_l$ represents sublattice $A$ or $B$ degree of freedom in layer index $l = \pm 1$\Ch{, $\boldsymbol{k}_{\text{mic}} = \boldsymbol{k} + (\boldsymbol{K}_{\xi}^{+1} + \boldsymbol{K}_{\xi}^{-1})/2 + \boldsymbol{\hat{x}} \sqrt{3} \xi / 2 | \boldsymbol{K}_{\xi}^{+1} - \boldsymbol{K}_{\xi}^{-1}|$ is the microscopic momentum of the electrons, $\boldsymbol{k}$ is the mini Brillouin zone momentum, and 
\begin{equation}
    \boldsymbol{K}^{l}_\xi = - \xi \frac{4\pi}{3a} R(-l\theta / 2) \boldsymbol{\hat{x}}
\end{equation}
for layer $l = \pm 1$ and $a = 0.246 \ \mathrm{nm}$}. In Eq. \ref{eq:continuum-hamiltonian}, $H_l^{\xi}$ are the monolayer graphene Hamiltonians, which, in close vicinity of the $\xi = \pm 1$ valleys, resemble Dirac cones:
\begin{equation}
    H_l^{\Ch{\xi}}\Ch{(\boldsymbol{k})} = -\hbar v_F^{\text{ml}} \left[ R(l\theta / 2)\Ch{(\boldsymbol{k}_{\text{mic}} - \boldsymbol{K}_{\xi}^{l})}  \right] \cdot (\xi \sigma_x, \sigma_y)
\end{equation}
where $R(\varphi)$ is the $2\times 2$ rotation matrix, and $v_F^{\text{ml}}$ is the monolayer Graphene Fermi velocity. The interlayer coupling is
\begin{align}
\begin{split}
    U_{\Ch{\xi}} = \begin{pmatrix}
        u & u' \\
        u' & u
    \end{pmatrix} &+ \begin{pmatrix}
        u & u' \nu^{-\xi} \\
        u' \nu^{\xi} & u
    \end{pmatrix} e^{i\xi \boldsymbol{G}_1 \cdot \boldsymbol{r}} \\
    &+ \begin{pmatrix}
        u & u' \nu^{\xi} \\
        u' \nu^{-\xi} & u
    \end{pmatrix} e^{i\xi (\boldsymbol{G}_2 + \boldsymbol{G}_3) \cdot \boldsymbol{r}}
\end{split}
\end{align}
Using minimal coupling, we obtain time-dependent monolayer graphene Hamiltonians, with Fourier transform
\begin{align} \label{eq:fourier-layer-h}
\begin{split}
    H_l^{\Ch{\xi}(n)}\Ch{(\boldsymbol{k})} = -\hbar v & \Big\{ R(l\theta / 2)\Big(\Ch{[\boldsymbol{k}_{\text{mic}} - \boldsymbol{K}_{\xi}^{(l)}] \delta_{n, 0}} \\
    &+ \frac{e}{\hbar} \frac{1}{2} \mathcal{E} [(\delta_{n, 1} + \delta_{n,-1}) \hat{\boldsymbol{y}} \\
    &- i (\delta_{n, -1} - \delta_{n,1})\hat{\boldsymbol{x}}] \Big)  \Big\} \cdot (\xi \sigma_x, \sigma_y).
\end{split}
\end{align}
Then,
\begin{equation}
    H_{\xi}^{(n)} = \begin{pmatrix}
        H_1^{\Ch{\xi}(n)}\Ch{(\boldsymbol{k})} & U^{\dagger}_{\Ch{\xi}} \delta_{n,0} \\
        U_{\Ch{\xi}} \delta_{n,0} & H_2^{\Ch{\xi}(n)}\Ch{(\boldsymbol{k})}
    \end{pmatrix}
\end{equation}
is the Fourier transform of the continuum model Hamiltonian. For the continuum model, we truncate the Floquet Hamiltonian (Eq. \ref{eq:continuum-hamiltonian}) to $-6 \leq m \leq 6$.

Upon diagonalizing the Floquet Hamiltonian, we obtain a large number of Floquet states per energy interval $[-\hbar \Omega/2, \hbar\Omega/2]$. We select two states per $\boldsymbol{k}$-point whose spectral weights $A^{0}_{\alpha}(\boldsymbol{k}) = |\langle \phi^{0}_{\boldsymbol{k}\alpha} | \phi^{0}_{\boldsymbol{k}\alpha} \rangle|^2$ are large (which makes their contribution to the Floquet-Boltzmann equation most important, see Sec. X). 

\subsection{Quasienergy Bands}
In Sec. \ref{sec:choice}, we provide and motivate the choices of physical parameters that we use in the main text. In Fig. \ref{fig:fig_bands}, we preview the quasienergy bands for our choice of toy and continuum model parameters.

\section{Choice of Physical Parameters} \label{sec:choice}
First, we present the physical parameters we use for the electronic Hamiltonian in the TBG continuum model (see Sec. \ref{sec:cont-model-hamiltonian} for the Hamiltonian). We consider the non-interacting continuum model \cite{macdonald, continuum-koshino} at a near-magic twist angle of $\theta = 1.13^{\circ}$. The bandwidth of the central bands at this angle is $W \approx 5 \ \mathrm{meV}$, and a perturbative expansion of the Hamiltonian around the Brillouin zone Dirac points \cite{macdonald} estimates the Fermi velocity as 
\begin{equation} \label{eq:v_f-nomin}
    v_F(\theta) =  v_F^{\text{ml}} ({1-3\beta^2})/({1+3\beta^2 (1 + \eta^2)}) ,
\end{equation}
where $\beta = u'/(\hbar k_{\theta} v_F^{\text{ml}})$ and $\eta = u/u'$ with $v_F^{\text{ml}} = 8 \times 10^{5} \ \mathrm{m/s}$, $k_\theta = 4 \pi / (3L_M)$, $u = 0.0797 \ \mathrm{eV}$, and $u' = 0.0975 \ \mathrm{eV}$ \cite{macdonald,continuum-koshino}. Eq. \ref{eq:v_f-nomin} predicts that the Fermi velocity at the chosen twist angle is $v_F = 27 \ \mathrm{km/s}$. However, the derivation of Eq. \ref{eq:v_f-nomin} approximates that $\Ch{H_l^{\xi}}$ is roughly $\theta$-independent and tends to overestimate $v_F$ (see Fig. 4 inset in \cite{macdonald}). We can obtain a better estimate by numerically calculating the Fermi velocity along the path $K$-$M$ in $\boldsymbol{k}$-space of the $\nu = +1$ band in the $\xi = +1$ valley. (This is the direction of maximum Fermi velocity.) The estimate yields $v_F = 17.5  \ \mathrm{km/s}$, and we hereafter use this value. In our Floquet Hamiltonian, we use a laser angular-frequency of $\Omega \approx W/\hbar \approx 5 \ \mathrm{meV}/\hbar$.

Second, we present the parameters we use for the electronic Hamiltonian of the TBG two-band toy tight binding model (see Sec. \ref{sec:toy-model-details} for the Hamiltonian). We choose our toy model Fermi velocity, frequency, and twist angle to roughly match those of the continuum model. Specifically, we use a twist angle of $\theta = 1.13^{\circ}$ and choose $W = 3.1 \ \mathrm{meV}$ so that the Fermi velocity $v_F = WL_M / (2 \sqrt{3} \hbar) = 17 \ \mathrm{km/s}$ roughly matches that of the continuum model at the same angle. In the toy model Floquet Hamiltonian, we choose $\Omega \approx 5 \ \mathrm{meV}/\hbar$. 

Third, we discuss the parameters we use for the TBG phonons. For both the continuum and toy models, we consider phonons speeds in the range of $c_{\text{ph}} \in [17.9 \ \mathrm{km/s}, 19.4 \ \mathrm{km/s}]$. In the toy model, $v_{\text{eff}}^0 = 18.9 \ \mathrm{km/s}$, and, in the continuum model, $v_{\text{eff}}^0 = 19.5 \ \mathrm{km/s}$, so the range of $c_{\text{ph}}$ we choose covers the regime $c_{\text{ph}} < v_{\text{eff}}^0$, in which the drive induces the opposite regime $c_{\text{ph}} > v_{\text{eff}}(\mathcal{E})$ when $\mathcal{E} > \mathcal{E}^*$. We also use the same phonon bath temperature of $T_{\text{ph}} = 1 \ \mathrm{K}$ for the toy and continuum model calculations. 

Please see Sec. \ref{sec:monkhorst-pack-sec} for details of the numerical $\boldsymbol{k}$-point grid \Ch{and Sec. \ref{sec:form-factor} for details of the toy model form factor.}

\section{Anomalous Hall Conductivity Calculations for the Continuum Model} \label{sec:continuum}
In this section, we repeat the calculations in the main text on the TBG continuum model \cite{continuum-koshino,macdonald}. We consider the non-interacting limit, setting $\epsilon \to \infty$ so that $I^{\text{el-el}}_{\boldsymbol{k}\alpha} = 0$. 
\begin{figure}[b]
    \centering
    \includegraphics[width=\linewidth]{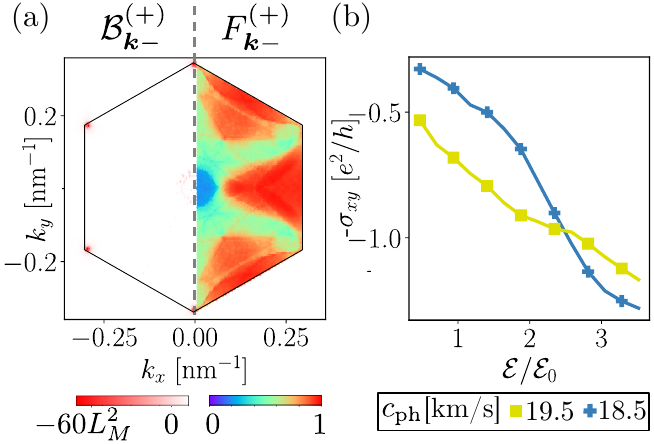}
    \caption{(a) Left: the steady-state occupation of the \Ch{lower} Floquet band in valley $\xi = +1$ of the continuum model \cite{continuum-koshino,macdonald}. Right: the Berry curvature of the same band, which peaks near the Dirac points and the resonance ring. (b) The anomalous Hall conductivity $\sigma_{xy}$ as a function of drive strength $\mathcal{E}$.}
    \label{fig:fig_continuum}
\end{figure}

First, we discuss differences in the bandstructure and topology at valleys $\xi = +$ and $\xi = -$. The circularly polarized laser opens a gap at the Dirac points, $\Delta_K$, effectively adding a mass term $\xi \Delta_K \sigma_z$ to the Hamiltonian (see Sec. \ref{sec:cont-model-hamiltonian} and \cite{Katz2020OpticallyGraphene} for a derivation) in the vicinity of the Dirac points. Because the sign of the mass term depends on $\xi$, the $\xi = \pm 1$ superlattice valley contributions to $\sigma_{xy}$ do not trivially cancel to zero. In fact, in reciprocal space, the Berry curvature and occupations near $\xi = +1$ are simple $\pi /3$ rotations of those in $\xi = -1$, so
\begin{equation}
    \sigma_{xy} = \frac{4e^2}{h} \sum_{\alpha=\pm}  \int_{\text{MBZ}} \frac{d^2\boldsymbol{k}}{(2\pi)^2} \mathcal{B}_{\boldsymbol{k}\alpha}^{(+1)} F_{\boldsymbol{k}\alpha}^{(+1)}.
\end{equation}
In Fig. \ref{fig:fig_continuum}, we show the steady-state and $\sigma_{xy}$ for the continuum model calculation. Note that we use the \Ch{full form factor $\mathcal{W}^{\xi\nu'\nu}_{\boldsymbol{k},\boldsymbol{q}} =\langle \xi \nu' \boldsymbol{k} + \boldsymbol{q} |\xi \nu \boldsymbol{k} \rangle$ as calculated from the continuum model wavefunctions (see Sec. \ref{sec:form-factor}).}

\section{Direct Variation of the Phonon Speed $c_{\mathrm{ph}}$}
\begin{figure}[b]
    \centering
    \includegraphics[width=0.85\linewidth]{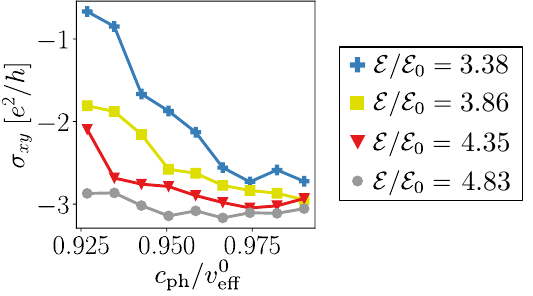}
    \caption{Anomalous Hall conductivity of the toy model as a function of the ratio $c_{\text{ph}} / v_{\text{eff}}^0$ for three different drive field strengths $\mathcal{E}/\mathcal{E}_0$. The same electron-phonon decoupling process is visible as $\sigma_{xy}$ plateaus.}
    \label{fig:fig_vary_c_s_conductivity}
\end{figure}

Throughout the main text, we use the drive strength $\mathcal{E}$ to control electron speeds. We could achieve similar results by keeping $\mathcal{E}$ fixed and varying $c_{\text{ph}}$ instead. Fig. \ref{fig:fig_vary_c_s_conductivity} shows the variation of $\sigma_{xy}$ as a function of $c_{\text{ph}}$. The curves resemble the dependence of $\sigma_{xy}$ on $\mathcal{E}$ in the main text (see, for e.g., Fig. 1(b)). 

{
\color{black}

\section{Symmetries and Berry Curvature Distributions of the Toy and Continuum Models}
In this section, we compare the symmetries of the toy and continuum models with and without the drive, and we show that the Berry curvature distributions of the models with the drive are consistent near the Dirac points.

\subsection{Symmetries} \label{sec:symmetries}
The symmetries of the undriven BM continuum model are $C_{2z} T = \sigma_x \mathcal{K}$, $C_3 = e^{i2\pi/3 \sigma_z}$, and $C_{2x} = \sigma_x \tau_x$, where $\tau_{i}$ and $\sigma_{i}$ are in the layer and sublattice degrees of freedom, $i = 0, x, y, z$, and $\mathcal{K}$ is conjugation. At charge neutrality, undriven TBG also has an emergent, approximate unitary particle-hole symmetry $P = i \tau_y$ and anti-unitary particle-hole symmetry $\mathcal{P} = P C_{2z} T$ within each superlattice valley, which ensures ${P} H_{\xi}({\boldsymbol{k}}) {P}^{-1} \approx -H_{\xi}(-{\boldsymbol{k}})$ and $\mathcal{P} H_{\xi}({\boldsymbol{k}}) \mathcal{P}^{-1} \approx -H_{\xi}(-{\boldsymbol{k}})$ (see Eq. \ref{eq:continuum-hamiltonian} for the definition of $H_{\xi}({\boldsymbol{k}})$) \cite{PhysRevB.103.205413, PhysRevB.103.205412, PhysRevLett.123.036401}. 

We now discuss the effect of the drive on the symmetries of the BM model. The drive induces a dynamical Haldane mass term $\Delta_K \xi \sigma_z$ (see Sec. \ref{sec:berry-comp}). It also resonantly-couples states around the resonance ring, where the effective Hamiltonian is  
\begin{equation}
    H_R^{\xi}(\boldsymbol{k}) = V_R(\boldsymbol{k}) | \xi - \boldsymbol{k} \rangle \langle \xi + \boldsymbol{k}|+ V_R^*(\boldsymbol{k}) | \xi + \boldsymbol{k} \rangle \langle \xi - \boldsymbol{k}|
\end{equation}
as derived from degenerate perturbation theory, where $V_R(\boldsymbol{k}) = \langle \xi - \boldsymbol{k} |H_{\xi}^{(1)}(\boldsymbol{k})| \xi + \boldsymbol{k} \rangle$ and $| \xi \pm \boldsymbol{k} \rangle$ are the undriven single-particle Bloch states (see the main text for the definition) \cite{KarniThroughMaterials}. We can see that the drive breaks $C_{2z}T$ symmetry by opening the Haldane gap. The drive also breaks $P$ symmetry because the drive-induced Haldane mass term $\Delta_K \xi \sigma_z$ commutes with $P$. However, the drive preserves the anti-unitary $\mathcal{P}$ symmetry. One can see this by first noting that $(C_{2z}T) \sigma_z (C_{2z}T)^{-1} = -\sigma_z$, so $\mathcal{P} \Delta_K \xi \sigma_z \mathcal{P}^{-1} = -\Delta_K \xi \sigma_z$. Secondly, since $\mathcal{P} | \xi -, \pm \boldsymbol{k} \rangle = | \xi +, \mp \boldsymbol{k} \rangle$, \begin{align}
\begin{split}
    \mathcal{P} H_R(\boldsymbol{k}) \mathcal{P}^{-1}& = V_R(\boldsymbol{k}) | \xi +,- \boldsymbol{k} \rangle \langle \xi -,- \boldsymbol{k}| \\
    &+ V_R^*(\boldsymbol{k}) | \xi -,- \boldsymbol{k} \rangle \langle \xi +,- \boldsymbol{k}|.
\end{split}
\end{align}
Noting that $\mathcal{P} H_{\xi}^{(1)}(\boldsymbol{k}) \mathcal{P}^{-1} = - H_{\xi}^{(-1)}(-\boldsymbol{k})$ and $| \xi + \boldsymbol{k} \rangle = \mathcal{P}^{-1} | \xi -,- \boldsymbol{k} \rangle$, we find that 
\begin{equation}
    V_R(\boldsymbol{k}) = - \langle \xi +,- \boldsymbol{k} |H_{\xi}^{(-1)}(-\boldsymbol{k})| \xi -,- \boldsymbol{k} \rangle = -V_R^*(-\boldsymbol{k}).
\end{equation}
Therefore, $\mathcal{P} H_R(\boldsymbol{k}) \mathcal{P}^{-1} = - H_R(-\boldsymbol{k})$, and the Hamiltonian is also particle-hole symmetric along the resonantly-coupled states. Lastly, the drive preserves $C_3$ symmetry and $C_{2x}$ symmetry, which one can see by noting the following:
\begin{equation}
    C_3 H_{\xi}(\boldsymbol{k}) C_3^{-1} = H_{\xi}(\boldsymbol{k}), \ C_{2x} H_{\xi}(\boldsymbol{k}) C_{2x}^{-1} = H_{-\xi}(\boldsymbol{k}),
\end{equation}
\begin{equation}
    C_3 \Delta_K \xi \sigma_z C_3^{-1} = \Delta_K \xi \sigma_z, \ C_{2x} \Delta_K \xi \sigma_z C_{2x}^{-1} = -\Delta_K \xi \sigma_z,
\end{equation}
and 
\begin{equation}
    C_3 H_R^{\xi}(\boldsymbol{k}) C_3^{-1} = H_R^{\xi}(\boldsymbol{k}), \ C_{2x} H_R^{\xi}(\boldsymbol{k}) C_{2x}^{-1} = H_R^{-\xi}(\boldsymbol{k}).
\end{equation}
Thus, the Hamiltonian near the Dirac points and resonantly-coupled states respect the $C_3$ and $C_{2x}$ symmetries.

Now, we discuss the symmetries of the toy model. In the undriven limit, the toy model has exact particle-hole symmetry $P = \sigma_z \mathcal{K}$. It also has the symmetries $C_{2z}T = \sigma_x \mathcal{K}$, $C = \sigma_z$ (sublattice/chiral symmetry), and $C_3 = e^{i2\pi/3\sigma_z}$. The drive opens a Haldane mass gap $\Delta_K \xi_{\text{MBZ}} \sigma_z$, where $\xi_{\text{MBZ}} = 1$ ($-1$) for the mini Brillouin zone $K$ ($K'$) point. We can now see that the drive breaks $C_{2z}T$, $C$, and $T$ symmetry via the Haldane mass term, while preserving $P = CT$ symmetry. One can see that $P$ is preserved by noting that $P\Delta_K \xi_{\text{MBZ}} \sigma_z P^{-1}= - \Delta_K \xi_{\text{MBZ}} \sigma_z$ since conjugation $\mathcal{K}$ inverts the momentum and hence the sign of the mass term. Similar arguments as the continuum model case can be made to show that the Hamiltonian near the resonance ring respects $P$.

Importantly, the drive preserves the emergent particle-hole symmetry in the continuum model while preserving the exact particle-hole symmetry in the toy model. As we note in the phenomenological analysis section of the main text, the emergent particle-hole symmetry ensures that the electron and hole scattering rates in the UFB and LFB are similar. Secondly, the drive breaks $C_{2z}T$ symmetry in both models by opening a Haldane gap. In Sec. \ref{sec:berry-comp}, we show that the Haldane gap ensures the Berry curvature distributions of the models near the Dirac points are consistent.

\subsection{Berry Curvature}
\label{sec:berry-comp}
The tunable conductivity $\sigma_{xy}$ relies only on the large Berry curvature and electron-phonon scattering bottlenecks near the Dirac points of the mini Brillouin zone. In this section, we detail how the Berry curvature distributions for the toy and continuum models are consistent near the Dirac points, as the numerical calculations of Berry curvature demonstrate in Fig. \ref{fig:fig_compare_berry}. We now prove the agreement analytically. The Hamiltonian for the mini Brillouin zone Dirac cone in the toy model is given by 
\begin{equation}
    H_{\text{Dirac}}^{\text{toy}}(\boldsymbol{q}) = \hbar v_F \boldsymbol{q} \cdot ( \xi_{\text{MBZ}}\sigma_x, \sigma_y),
\end{equation}
where $\boldsymbol{q}$ is the momentum measured from the $K$ or $K'$ point, $v_F$ is the Fermi velocity of TBG, and $\xi_{\text{MBZ}} = +$ for the $K$ point and $\xi_{\text{MBZ}} = -$ for the $K'$ point in the mini Brillouin zone. The corresponding Hamiltonian for the continuum model is
\begin{equation}
    H_{\text{Dirac}}^{\text{cont}}(\boldsymbol{q}) = \hbar v_F \boldsymbol{q} \cdot (\xi\sigma_x,  \sigma_y)
\end{equation}
where $\xi$ is the superlattice valley index. Upon applying minimal coupling and the Van-Vleck perturbative expansion (see Sec. \ref{sec:gap-sizes} for details), one finds the effective Floquet Hamiltonians
\begin{equation}
     H_{\text{Dirac,eff}}^{\text{toy}}(\boldsymbol{q}) = \hbar v_F \boldsymbol{q} \cdot (\xi_{\text{MBZ}} \sigma_x, \sigma_y) + \xi_{\text{MBZ}} \Delta_K \sigma_{z}
\end{equation}
\begin{equation} \label{eq:cont-eff-dirac}
     H_{\text{Dirac,eff}}^{\text{cont}}(\boldsymbol{q}) = \hbar v_F \boldsymbol{q} \cdot (\xi \sigma_x,  \sigma_y) + \xi \Delta_K \sigma_{z}.
\end{equation}
We note that in both models, the Berry curvature does not alternate signs between the $K$ and $K'$ points in mini Brillouin zone. Additionally, Eq. \ref{eq:cont-eff-dirac} shows that the drive breaks time-reversal symmetry between the superlattice valleys $\xi = \pm 1$ in the BM model, permitting nonzero $\sigma_{xy}$ when contributions to the conductivity from both superlattice valleys are combined.

\begin{figure}[b]
    \centering
    \includegraphics[width=\linewidth]{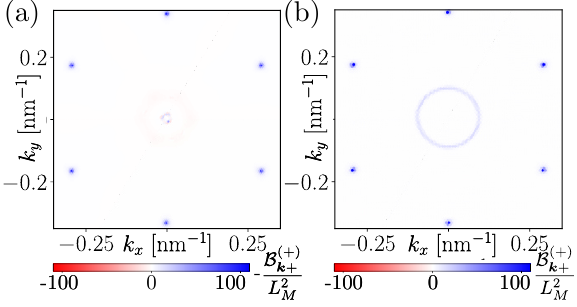}
    \caption{\Ch{Comparing the Berry curvature distribution in the upper Floquet band in the (a) continuum model and (b) toy model at a drive amplitude of $\mathcal{E}/\mathcal{E}_0 \approx 3.1$.}}
    \label{fig:fig_compare_berry}
\end{figure}

\section{Full Phenomenological Model} \label{sec:full-phenom}
In this section, we derive a detailed phenomenological model that qualitatively reproduces the dependence of $\sigma_{xy}$ on $\mathcal{E}$ and the effect of interactions presented in the main text. 

Let us begin by adding more details to the phenomenological model for the $K$-point occupation. We write
\begin{align} \label{eq:bigger-f-kp}
\begin{split}
    F_{K+}^{(\xi)} = [&\mathcal{R}_{\text{in}} F_{+} + \Gamma^{\text{el}}_{\text{in}} + \Gamma^{\text{el,FU}}_{\text{in}} +\mathcal{R}^{\text{FU}}_{\text{in}}F_{-}] / \\
    & [\mathcal{R}_{\text{in}} F_{+} + \mathcal{R}_{\text{out}}(1-F_{-}) \\ 
    &+ \Gamma^{\text{el}}_{\text{in}} + \Gamma^{\text{el,FU}}_{\text{in}}  +\Gamma^{\text{el}}_{\text{out}} + \Gamma^{\text{el,FU}}_{\text{out}} \\
    &+\mathcal{R}^{\text{FU}}_{\text{in}}F_{-} + \mathcal{R}^{\text{FU}}_{\text{out}}(1-F_{+})],
\end{split}
\end{align}
where $\Gamma^{\text{el}}_{\text{in/out}}$, $\Gamma^{\text{el, FU}}_{\text{in/out}}$ and $\mathcal{R}^{\text{FU}}_{\text{in/out}}$ are the non-FU electron-electron, FU electron-electron, and FU electron-phonon scattering rates, respectively. Here, $F_{\alpha}$ is the average occupation of Floquet band $\alpha$ outside the resonance ring, and $F_{+} = 1-F_{-}$. We drop the superscript $(\xi)$ on the occupations for simplicity and work within a single superlattice valley. Let us now make the following approximations and definitions:
\begin{equation}
     \Gamma^{\text{el}}_{\text{in}} \approx \Gamma^{\text{el}}_{\text{out}} \equiv \Gamma , \ \text{and}  \ \Gamma^{\text{el,FU}}_{\text{in}} \approx \Gamma^{\text{el,FU}}_{\text{out}} \approx S \Gamma,
\end{equation}
where FU processes are suppressed by a factor of $S \equiv (V/\hbar\Omega)^2$, with $V \approx v_F e \mathcal{E}/\Omega$. For phonon transitions, let us make the following definitions:
\begin{equation}
    \mathcal{R}_{\text{out}} \equiv \mathcal{R}, \ \text{and} \ \mathcal{R}_{\text{in}} = r \mathcal{R},
\end{equation}
where $r \equiv \mathcal{R}_{\text{in}} / \mathcal{R}_{\text{out}}$. Let us also define $r^{\text{FU}} \equiv \mathcal{R}_{\text{in}}^{\text{FU}} / \mathcal{R}_{\text{out}}^{\text{FU}}$, and approximate $\mathcal{R}_{\text{out}}^{\text{FU}}\approx S a^{\text{FU}} \mathcal{R}$, where the factor $a^{\text{FU}} > 1$ accounts for the fact that the phase space area of states connected to the UFB $K$ point by FU processes is much larger than area of states connected to the UFB $K$ point by non-FU processes. We therefore obtain
\begin{equation}
     \mathcal{R}_{\text{out}}^{\text{FU}} = S  a^{\text{FU}}\mathcal{R}, \ \text{and} \ \mathcal{R}_{\text{in}}^{\text{FU}} = Sa^{\text{FU}} r^{\text{FU}}\mathcal{R}.
\end{equation}
Now, Eq. \ref{eq:bigger-f-kp} reduces to
\begin{equation}
    F_{K+}^{(\xi)} = \frac{r F_{+}\mathcal{R} + S  F_{-}a^{\text{FU}} r^{\text{FU}}\mathcal{R} + (1+S) \Gamma}{(1+r) F_{+} \mathcal{R} + S(1+r^{\text{FU}}) F_{-} a^{\text{FU}}\mathcal{R} + 2(1+S) \Gamma}.
\end{equation}
Let us further define $x \equiv \Gamma/\mathcal{R}$ as a ratio of electron-electron to electron-phonon scattering rates and use $F_- \approx 1-F_+$ (ensured by emergent particle-hole symmetry, see Sec. \ref{sec:symmetries}) to obtain
\begin{equation} \label{eq:full-f-k-p}
    F_{K+}^{(\xi)} = \frac{r F_{+} + S  (1-F_{+})a^{\text{FU}} r^{\text{FU}} + (1+S) x}{(1+r) F_{+} + S(1+r^{\text{FU}}) (1-F_{+})a^{\text{FU}}  + 2(1+S) x}.
\end{equation}
We will determine the dependence of $F_{K+}^{(\xi)}$ on interaction strength $x$ at strong and weak drive amplitudes.

We now derive the phenomenological equation for $F_{\alpha}$. The rate equation is roughly
\begin{align}
\begin{split}
    \dot{F}_{+} \approx &\Lambda_{\text{in}} F_- (1-F_+) - \Lambda_{\text{out}} F_+(1-F_-) \\
    &+\Gamma (1+S) (1-F_+) - \Gamma (1+S) F_+
\end{split}
\end{align}
where $\Lambda_{\text{in}}$ and $\Lambda_{\text{out}}$ are electron-phonon scattering rates into and out of the UFB. Note that $\Lambda_{\text{in}} \approx S \Lambda_{\text{out}}$ since scattering processes described by $\Lambda_{\text{in}}$ are FU processes. We also approximate $\Lambda_{\text{out}} \approx f_b \mathcal{R}$, with a factor $f_b > 1$, because we expect $\mathcal{R}$, which is suppressed by strong electron-phonon scattering bottlenecks near the $K$ point, to be smaller than $\Lambda_{\text{out}}$, the total scattering rate into the UFB. We can then find the steady-state solution $F_+$ in terms of $x$ and substitute the results into Eq. \ref{eq:full-f-k-p}.

\begin{figure}
    \centering
    \includegraphics[width=0.9\linewidth]{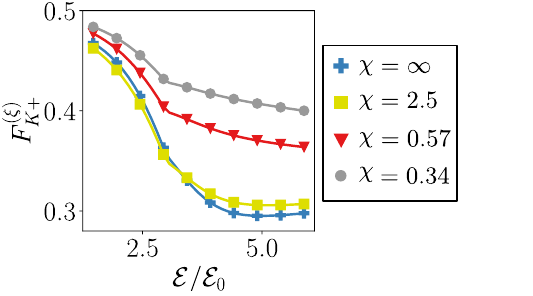}
    \caption{\Ch{The occupation $F^{(\xi)}_{K+}$ as predicted by the phenomenological model in Sec. \ref{sec:full-phenom} for different ratios $\chi \equiv \tau^{\text{el}}_{K} / \tau^{\text{ph}}_{K}$. Note that for large $\chi$, the occupation is \textit{lower} than the non-interacting case at weak drive amplitudes, an effect of reduced Pauli blocking in the electron-phonon interactions.}}
    \label{fig:fig_fullphenom}
\end{figure}

In Fig. \ref{fig:fig_fullphenom}, we show the occupation $F_{K+}^{(\xi)}$ as a function of $\mathcal{E}$ for different values of $\chi = \tau^{\text{el}}_{K} / \tau^{\text{ph}}_{K} \approx x^{-1} F_+|_{\mathcal{E}=\mathcal{E}^*}$ (c.f. $1/\tau^{\text{ph}}_{K} \sim \mathcal{R} F_+|_{\mathcal{E}=\mathcal{E}^*}$ and $1/\tau^{\text{el}}_{K} \sim \Gamma$) where $F_+|_{\mathcal{E}=\mathcal{E}^*}$ is $F_+$ evaluated at the drive ampltiude $\mathcal{E}^*$, which we choose to be $\mathcal{E}^* = 2.5 \mathcal{E}_0$. To generate the figure, we choose $f_b = 3$ and $a^{\text{FU}} = 28$. Note that $a^{\text{FU}}$ estimates that the total area of momentum states connected to $\mathcal{A}_K$ by electron-phonon FU processes covers roughly $1/6$ of the Brillouin zone. We write the following $\mathcal{E}$-dependent phenomenological equations for the ratios $r$ and $r^{\text{FU}}$ that capture very roughly their dependence on $\mathcal{E}$, inspired by Fig. 4(b) in the main text. First, we approximate
\begin{equation}
    r \approx \mathrm{max}\left[\frac{2}{e^{(\mathcal{E}-\mathcal{E}^*)/(0.8 \mathcal{E}_0)}+1} - 1, 0\right].
\end{equation}
Note $r \sim 1$ for $\mathcal{E} \ll \mathcal{E}^*$ and $r = 0$ for $\mathcal{E} > \mathcal{E}^*$, capturing the behavior shown in Fig. 4(b) in the main text. Second, we choose
\begin{equation}
    r^{\text{FU}} \approx \frac{1/2}{e^{(\mathcal{E}-\mathcal{E}^*)/(0.5 \mathcal{E}_0)}+1} + \frac{1}{2}.
\end{equation}
Here, $r^{\text{FU}} \sim 1$ for $\mathcal{E} \ll \mathcal{E}^*$ and decreases with $\mathcal{E}$, but never reaches $0$ (since the electrons are never decoupled from electron-phonon FU processes). 

Two features are notable in Fig. \ref{fig:fig_fullphenom}. First, the occupation decreases as a function of interaction strength for weak interactions (large $\chi$) and weak drive amplitudes, a result of interactions reducing Pauli blocking of the phonon processes from patch $\mathcal{S}_{\text{in}}$ to $\mathcal{S}_K$ and from $\mathcal{S}_K$ to $\mathcal{S}_{\text{out}}$ (i.e., increasing $F_{\text{in}}^{(\xi)}$ in Eq. 6 of the main text, correspondingly suppressing $F_{K+}^{(\xi)}$). When interactions are strong (small $\chi$), $\Gamma^{\text{el}}_{\text{in/out}}$ dominates, and $F_{K+}^{(\xi)} \to 0.5$. Second, the occupation increases slowly for $\mathcal{E} > \mathcal{E}^*$, since FU processes strengthen as $(V/\hbar\Omega)^2$ grows with $\mathcal{E}$. Both of these behaviors are visible in Fig. 3(c) in the main text.
}

\section{Formal Definition, Numerical Evaluation, and Phenomenological Model of $\mathcal{A}_{\text{in}}$}
As described in the main text, a patch $\mathcal{S}_{\text{in}}$ shaped as an elliptical annulus (see Fig. 3(a)) with area $\mathcal{A}_{\text{in}}$ in momentum space vanishes as $\mathcal{E} \to \mathcal{E}^*$. Here, we provide a formal definition of $\mathcal{A}_{\text{in}}$ and explain how we estimate its dependence on $\mathcal{E}$ numerically and analytically. 

\subsection{Formal Definition}
Let us first define $\mathcal{A}_{\text{in}}$ formally. Consider a family of phonon cones centered throughout $\mathcal{S}_{K}$, the circular patch enclosing a $K$-point in the quasienergy spectrum (see Fig. 3(a)). Suppose that a subset of the phonon cones are centered throughout a small quasienergy window $d\varepsilon_{\boldsymbol{k}+}$. The $\boldsymbol{k}$-space area of states $d\mathcal{A}_{\text{in}}$ containing intersections of the cones with the upper Floquet band is
\begin{equation}
    d\mathcal{A}_{\text{in}} = d\varepsilon_{\boldsymbol{k}+}\sum_{s=\pm} \int d^2 \boldsymbol{k}' \  \delta(\varepsilon_{\boldsymbol{k}+} - \varepsilon_{\boldsymbol{k}'+} + s \hbar c_{\text{ph}} |\boldsymbol{k}'-\boldsymbol{k}|).
\end{equation}
Next, we integrate over $\varepsilon_{\boldsymbol{k}+}$ contained in $\mathcal{S}_{K}$ to obtain
\begin{equation} \label{eq:a-in-formal}
\begin{split}
    \mathcal{A}_{\text{in}} = & \int d\mathcal{A} = \int_{\boldsymbol{k} \in \mathcal{S}_{K}} d^2 \boldsymbol{k} \frac{1}{D(\varepsilon_{\boldsymbol{k}+})} \times  \\
    & \times \left[ \sum_{s=\pm} \int d^2 \boldsymbol{k}' \ \delta(\varepsilon_{\boldsymbol{k}+} - \varepsilon_{\boldsymbol{k}'+} + s \hbar c_{\text{ph}} |\boldsymbol{k}'-\boldsymbol{k}|)  \right],
\end{split}
\end{equation}
where 
\begin{equation}
    D(\varepsilon) =\sum_{\alpha} \int \frac{d^2\boldsymbol{k}}{(2\pi)^2} \delta(\varepsilon - \varepsilon_{\boldsymbol{k}\alpha})
\end{equation}
is the density of states in the quasienergy band structure. 
Exploiting the circular shape of $\mathcal{S}_{K}$,
\begin{equation}
    \int_{\boldsymbol{k} \in \mathcal{S}_{K}} d^2 \boldsymbol{k} \approx \int d^2{\boldsymbol{k}} \ \Theta(|\boldsymbol{k} - \boldsymbol{K}| - k_p)
\end{equation}
where $k_p$ is the radius of the circular area $\mathcal{A}_K$ of $\mathcal{S}_K$. Lastly, we calculate an approximate expression for $k_p$, the radius of $\mathcal{A}_{K}$. In the vicinity of the Dirac cone, the Hamiltonian is 
\begin{equation}
    H_K(\boldsymbol{k},t) = \boldsymbol{d} \cdot \boldsymbol{\sigma},
\end{equation}
where $\boldsymbol{d} = \hbar v_F \xi k_x \boldsymbol{\hat{x}} + \hbar v_F k_y \boldsymbol{\hat{y}} + \xi \Delta_K \mathcal{E}^2 \boldsymbol{\hat{z}}$. (See Sec. \ref{sec:gap-sizes} for a detailed derivation.) The $\boldsymbol{z}$-component of the Berry curvature is 
\begin{equation}
    \mathcal{B}_{{k}\alpha}^z = \alpha \frac{{d}_z}{2|\boldsymbol{d}|^3} = \alpha  \frac{\Delta_K}{[(\hbar v_F k)^2 + \Delta_K^2]^{3/2}}
\end{equation}
where $d_z = \xi \Delta_K$ \Ch{and $\alpha = \pm$}. At the half-maximum, $\mathcal{B}^z_{k_p\alpha} = 0.5 \mathcal{B}^z_{0\alpha}$, so
\begin{equation}
    k_p = (2^{2/3} -1)^{1/2} \frac{\Delta_K}{\hbar v_F}.
\end{equation}

\subsection{Numerical Estimate}
To generate the values of $\mathcal{A}_{\text{in}}$ we present in Fig. 3(b), we evaluate the integrals in Eq. \ref{eq:a-in-formal} on a finite-sized grid of $\boldsymbol{k}$-points, smearing the step function by replacing $\Theta(|\boldsymbol{k} - \boldsymbol{K}| - k_p) \to [e^{(|\boldsymbol{k} - \boldsymbol{K}| - k_p) / \sigma_k} + 1]^{-1}$, where $\sigma_k = 2\pi / (L_M N)$ is the grid spacing between $\boldsymbol{k}$-points on an $N \times N$ Monkhorst-Pack grid (see Sec. \ref{sec:monkhorst-pack-sec}). Thus, we approximate
\begin{equation} 
\begin{split}
    \mathcal{A}_{\text{in}} \approx
    &\sum_{\boldsymbol{k}} [e^{(|\boldsymbol{k} - \boldsymbol{K}| - k_p) / \sigma_k} + 1]^{-1} \frac{1}{D(\varepsilon_{\boldsymbol{k}+})} \times  \\
    & \times \left[ \sum_{s=\pm} \sum_{\boldsymbol{k}'} \ \delta(\varepsilon_{\boldsymbol{k}+} - \varepsilon_{\boldsymbol{k}'+} + s \hbar c_{\text{ph}} |\boldsymbol{k}'-\boldsymbol{k}|)  \right].
\end{split}
\end{equation}
For more information on how we approximate the Dirac Delta function on the grid, please see Sec. \ref{sec:en-mom-conservation}. {
\color{black} Note that we numerically-estimate the $K$-point occupation in a similar way, by calculating 
\begin{equation}
    F^{(\xi)}_{K+} \approx \sum_{\boldsymbol{k}} [e^{(|\boldsymbol{k} - \boldsymbol{K}| - k_p) / \sigma_k} + 1]^{-1} F_{\boldsymbol{k}+}^{(\xi)}.
\end{equation}}

\subsection{Phenomenological Model}
\begin{figure}
    \centering    \includegraphics[width=\linewidth]{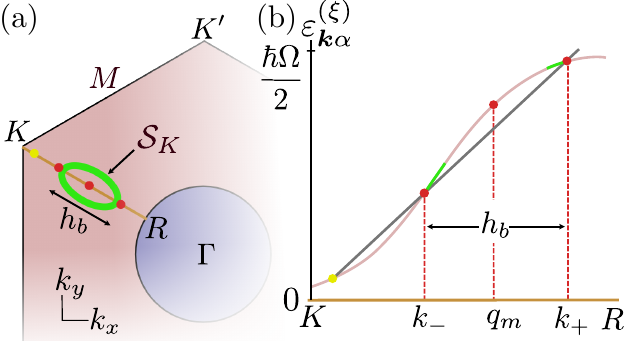}
    \caption{(a) The intersection $\mathcal{S}_K$ (Fig. 3(a)) as viewed on the Brillouin zone. The outer radius along the path $KR$ is $h_b(\mathcal{E})$. (b) Quasienergy (pink) along the path $KR$, with the phonon light cone (grey) that determines the \textit{outer} radius of $\mathcal{A}_{\text{in}}$. The intersections $k_+$ and $k_-$ between the cone and the upper Floquet band determines $h_b(\mathcal{E}) = k_+ - k_-$.}
    \label{fig:fig_functional_h}
\end{figure}

\begin{figure}
    \centering    \includegraphics[width=\linewidth]{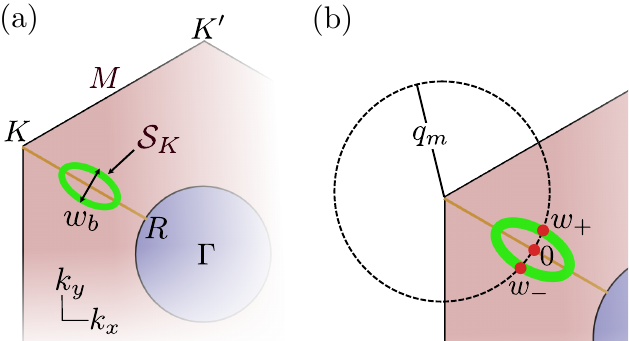}
    \caption{(a) Width of the intersection $\mathcal{S}_K$, $w_b(\mathcal{E})$. (b) Circular coordinate system with arc length $w$ (increasing counterclockwise) that we use to determine $w_b(\mathcal{E}) = w_+ - w_-$.}
    \label{fig:fig_functional_w}
\end{figure}

In this section, we prove that the intersection area $\mathcal{A}_{\text{in}} \propto \max(\mathcal{E}^* - \mathcal{E}, 0)$ as $\mathcal{E} \to \mathcal{E}^*$. The shape of $\mathcal{A}_{\text{in}}$ is an elliptical annulus as shown in Fig. 3(a). Let us use $h_b(\mathcal{E})$ and $w_b(\mathcal{E})$ respectively to denote the \textit{outer} major and minor axis radii of the elliptical annulus (see Figs. \ref{fig:fig_functional_h}(a) and \ref{fig:fig_functional_w}(a)). In the following sections, we begin by generating analytical estimates of $h_b(\mathcal{E})$ and $w_b(\mathcal{E})$.

\subsubsection{Estimate of $h_b$}

First, let us consider a slice of the upper Floquet band in $\boldsymbol{k}$-space from the $K$ to the resonance ring (R) along the direction of $h_b(\mathcal{E})$, as we show in Fig. \ref{fig:fig_functional_h}(b). Let us define a one-dimensional momentum component $q$ along the path $K$-$R$. We sketch a phonon light cone (grey) originating from a point (yellow) in $\mathcal{S}_{K}$ that determines the outer radius of $\mathcal{A}_{\text{in}}$. The phonon cone intersects with the quasienergy at points $k_+$ and $k_-$, and the outer radius of $\mathcal{A}_{\text{in}}$ is therefore $h_b(\mathcal{E}) = k_+ - k_-$. First, consider the undriven limit $\mathcal{E} = 0$, where the gaps $\Delta_R =0$ and $\Delta_K=0$. We choose some point $q_m$ such that $k_- < q_m < k_+$ and series expand the energy $E(q)$ of the undriven system around $q_m$:
\begin{align} \label{eq:taylor-series-quasi}
\begin{split}
    E(q) &\approx E(q_m) + E'(q_m) (q-q_m) + \frac{1}{2} E''(q_m)(q-q_m)^2\\
    &= a_2 q^2 + a_1 q + a_0,
\end{split}
\end{align}
where $a_2 = E''(q_m)/2$, $a_1 = E'(q_m) - E''(q_m) q_m$, and $a_0 =E(q_m) - E'(q_m) q_m + E''(q_m) q_m^2 / 2$. As we increase $\mathcal{E}$, the gaps $\Delta_K$ and $\Delta_R$ widen. Let us write the quasienergy in the vicinity of $q_m$ as
\begin{equation}
    \varepsilon(q) \approx f(\mathcal{E}) E(q) + \frac{\Delta_K}{2}
\end{equation}
where $f(\mathcal{E}) \leq 1$ is a scaling factor that decreases as $\mathcal{E}$ increases and accounts for band flattening due to $\Delta_K$ and $\Delta_R$. Let 
\begin{equation} \label{eq:factor-f}
    f^{-1} = 1-b_1 \tilde{\mathcal{E}} - b_2 \tilde{\mathcal{E}}^2,
\end{equation}
where $b_1 \geq 0$ and $b_2 \geq 0$ are constants dependent on the exact bandstructure (i.e., how the widening of $\Delta_K$ and $\Delta_R$ with $\mathcal{E}$ affects the bandstructure near $q_{m}$). The roots of the equation $E(q) = \Delta_K / 2 + \hbar c_{\text{ph}} q$ are $k_{\pm}$, and we may write the equation as
\begin{equation}
    a_2 q^2 + a_1 q + a_0 = f \hbar c_s q,
\end{equation}
from which we find that 
\begin{equation}
    h_b = k_+ - k_- = \sqrt{(a_1 - f\hbar c_s)^2 - 4a_2 a_0 }.
\end{equation}
Solving for $\mathcal{E}^*$ through the equation $h_b = 0$, and then series expanding the expression $(a_1 - f\hbar c_s)^2 - 4a_2 a_0$ in powers of small $\mathcal{E} - \mathcal{E}^*$, we find that $(a_1 - f\hbar c_s)^2 - 4a_2 a_0 \sim \mathcal{E}^* - \mathcal{E}$, so  $h_b \sim \sqrt{\mathcal{E}^* - \mathcal{E}}$.  

\subsubsection{Estimate of $w_b$}

To estimate $w_b$ (see Fig. \ref{fig:fig_functional_w}(a)), we define a circular coordinate system shown in Fig. \ref{fig:fig_functional_w}(b) whose origin is the $K$ point and arc length $w$ is zero along the $KR$ slice, increasing counterclockwise. The quasienergy $\varepsilon(w)$ along the circle perimeter varies with $w$; let us approximate
\begin{equation} \label{eq:delta_w}
    \varepsilon(w) \approx \hbar\Omega / 2 - (d_0 + d_2 w^2),
\end{equation}
using some fitting parameters $d_0$ and $d_2$. (We assume that $w = 0$ is at local maximum of $\epsilon(w)$, so there is no linear term in Eq. \ref{eq:delta_w}.) Roughly, $w_b = w_+ - w_-$, where we find $w_+$ and $w_-$ by finding the roots of the equation
\begin{equation}
	\hbar\Omega / 2 - \Delta(w) = f \hbar c_s q_m.
   \end{equation}
Here, once again, we use the factor $f$ in Eq. \ref{eq:factor-f} to account for band flattening as $\mathcal{E}$ increases from zero. So,
 \begin{equation}
 	w_b = w_+ - w_- = 2 \sqrt{(f\hbar c_s q_m + \hbar\Omega / 2 - d_0) / d_2}.
 \end{equation}
Solving for $\mathcal{E}^*$ by setting $w_b = 0$ and series expanding $f\hbar c_s q_m + \hbar\Omega / 2 - d_0$ in powers of $\mathcal{E}$, we find that $w_b \sim \sqrt{\mathcal{E}^* - \mathcal{E}}$. 

\subsubsection{Estimate of $\mathcal{A}_{\mathrm{in}}$}
In the limit $\mathcal{E} \to \mathcal{E}^*$, the elliptical annulus with finite thickness collapses into a filled ellipse. Thus, in the limit $\mathcal{E} \to \mathcal{E}^*$, we estimate that $\mathcal{A}_{\text{in}} = \pi h_b(\mathcal{E}) w_b(\mathcal{E}) \propto \max(\mathcal{E}^* - \mathcal{E}, 0)$.

\section{Predicting $\mathcal{E}^*$ for the Toy Model} \label{sec:prediction-ep-star-c-intra}

Here, we use the quasienergy dispersion of the toy model to predict $\mathcal{E}^*$. By writing an approximate, analytic expression for $v_{\text{eff}}(\mathcal{E})$ (see Eq. 6), we can find $\mathcal{E}^*$ using the relation $v_{\text{eff}}(\mathcal{E}^*) = c_{\text{ph}}$.
\begin{figure}
    \centering
    \includegraphics[width=\linewidth]{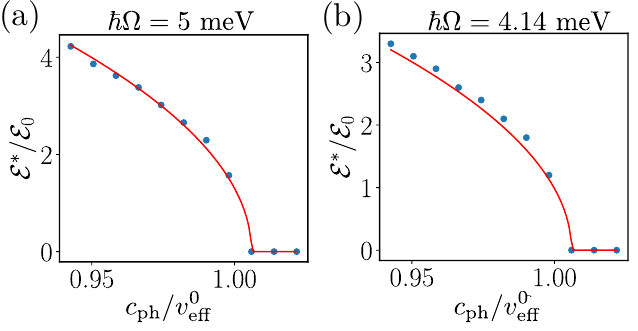}
    \caption{Comparing numerical evaluation of $\mathcal{E}^*$ (points) to an analytic fit to Eq. \ref{eq:supp-ep-star}. We use the same fitting parameters $f_2 = 0.778$, $f_1 = 0$, and $\delta(N) = 0.006$ for both panels.}
    \label{fig:fig_kink_locataion}
\end{figure}
From Eq. 6, $v_{\text{eff}}(\mathcal{E}) = ({\varepsilon_{\boldsymbol{k}^{*}+} - \varepsilon_{\boldsymbol{K}+}})/{|\boldsymbol{k}^{*}-\boldsymbol{K}|}$ for some appropriately-chosen $\boldsymbol{k}^*$ (dropping the superlattice valley index for notational simplicity). One can find numerically that $\boldsymbol{k}^*$ does not shift significantly with $\Omega$ or $\mathcal{E}$. We write an ansatz
\begin{equation}
    \varepsilon_{\boldsymbol{k}^*+} \approx \hbar v_{\text{eff}}^0 |\boldsymbol{k}^*- \boldsymbol{K}| - \frac{\hbar v_F}{L_M} \left(f_1' \tilde{\mathcal{E}} +  f_2' \tilde{\mathcal{E}}^2\right) \frac{|\boldsymbol{k}^*- \boldsymbol{K}|}{\Omega/(2v_{\text{eff}}^0)},
\end{equation}
where $f_1'$ and $f_2'$ are fitting constants dependent on the quasienergy bandstructure. Here, $\hbar v_F / L_M$ is the order of magnitude energy scale of the resonance ring gap $\Delta_R$. The dependence of $\varepsilon_{\boldsymbol{k}^*+}$ on $\mathcal{E}$ arises predominantly from $\Delta_R$. The dependence is stronger when $\boldsymbol{k}^*$ is close to the resonance ring, and we encode this behavior in the ratio ${|\boldsymbol{k}^*- \boldsymbol{K}|}/{\Omega/(2v_{\text{eff}}^0)}$, where $\Omega / (2v_{\text{eff}}^0)$ is the $\boldsymbol{k}$-space distance between the $K$ point and the resonance ring. Separately, we know that $\varepsilon_{\boldsymbol{K}+} = \Delta_K / 2$. We use Eq. 6 to infer
\begin{equation}
    v_{\text{eff}}^0(\mathcal{E}) = v_{\text{eff}}^0 - \frac{\Delta_K}{2\hbar |\boldsymbol{k^*}- \boldsymbol{K}|} - \frac{2\hbar v_F v_{\text{eff}}^0}{L_M \Omega} \left(f_1' \tilde{\mathcal{E}} +  f_2' \tilde{\mathcal{E}}^2\right).
\end{equation}
We know that $v_{\text{eff}}^0 \propto v_F$. We also assume that $|\boldsymbol{k^*}-\boldsymbol{K}|$ does not change significantly with $\mathcal{E}$, so it is independent of the drive and only dependent on the superlattice scale: $|\boldsymbol{k^*}-\boldsymbol{K}| \propto L_M^{-1}$. Thus, we can absorb some unknown coefficients into new coefficients $f_1''$ and $f_2''$ to obtain 
\begin{equation} \label{eq:approx-max-driven-velocity}
    v_{\text{eff}}^0(\mathcal{E}) = v_{\text{eff}}^0 - \frac{\hbar v_F^2}{L_M \Omega} \left(f_1''\tilde{\mathcal{E}} +  f_2'' \tilde{\mathcal{E}}^2\right).
\end{equation}
Upon solving for $\mathcal{\tilde{E}}^*$ from $c_s =v_{\text{eff}}(\mathcal{E}^*)$, we find that
\begin{equation} 
    \tilde{\mathcal{E}}^* \approx \sqrt{\frac{L_M \Omega}{3f_2 v_{F}}} \left(\sqrt{1-c_{\text{ph}}/v_{\text{eff}}^0 + f_1^2} - f_1 \right),
\end{equation}
where $f_1$ and $f_2$ are new, rescaled fitting constants. Using the relation $\tilde{\mathcal{E}} = e L_M \mathcal{E} / (\sqrt{3} \hbar\Omega)$, we find
\begin{equation} \label{eq:supp-ep-star}
    {\mathcal{E}}^* \approx {\frac{\hbar \Omega^{3/2}}{f_2 e L_M^{1/2}  v_{F}^{1/2}}} \left(\sqrt{1-c_{\text{ph}}/v_{\text{eff}}^0 + f_1^2} - f_1 \right),
\end{equation}
As $c_{\text{ph}} \to v_{\text{eff}}^0$, $\mathcal{E}^* \propto (1-c_{\text{ph}}/v_{\text{eff}}^0)^{\gamma}$ where $\gamma = 1$ ($1/2$) if $f_1 \neq 0$ ($= 0$). See Fig. \ref{fig:fig_kink_locataion} for a fit for two different frequencies $\Omega$.

Finite grid size effects on an $N \times N$ Monkhorst-Pack grid (see Sec. \ref{sec:monkhorst-pack-sec}) generate a small numerical error $\delta(N)$ that enters \ref{eq:supp-ep-star} as 
\begin{equation} \label{eq:supp_ep_star_finitegrid}
    {\mathcal{E}}^* \approx {\frac{\hbar L_M^{1/2} \Omega^{3/2}}{f_2 e L_M v_{F}^{1/2}}} \left(\sqrt{1-c_{\text{ph}}/v_{\text{eff}}^0 +\delta(N)+ f_1^2} - f_1 \right).
\end{equation}
To see this, let us consider the details of the finite-sized grid. We impose energy conservation through a broadened Dirac Delta function (see Sec. \ref{sec:en-mom-conservation}), which we model as a Gaussian function in energy with a tiny width
\begin{equation}
    \sqrt{2} \sigma \approx 0.1 \cdot \sqrt{2} \cdot \frac{W}{2N/3}.
\end{equation}
(We motivate the choice of the prefactor of $0.1$ in Sec. \ref{sec:en-mom-conservation}.) Since we avoid the high symmetry $K$ point in our grids, the $\boldsymbol{k}$-point with largest Berry curvature is, in fact, a point $\boldsymbol{k}_{\text{near}}$ point shifted away from $K$ by a small distance in momentum space of 
\begin{equation}
    |\delta \boldsymbol{k}| = |\boldsymbol{k}_{\text{near}} - \boldsymbol{K}| \approx \frac{1}{2} \frac{\Omega/(2\hbar v_{\text{eff}}^0)}{2N/3} = \frac{\Omega}{4v_{\text{eff}}^0(2N/3)}.
\end{equation}
This point is shifted in quasienergy by $\hbar v_F |\delta \boldsymbol{k}|$ relative to the actual $K$ point. We can account for both of these effects by shifting $\varepsilon_{\boldsymbol{K}+} \to \varepsilon_{\boldsymbol{K}+} +\delta \varepsilon$, with $\delta \varepsilon = \sqrt{2}\sigma + \hbar v_F |\delta \boldsymbol{k}|$ and solve $v_{\text{eff}}(\mathcal{E}^*) = c_{\text{ph}}$ to find Eq. \ref{eq:supp_ep_star_finitegrid} with $\delta(N) = \delta \varepsilon / (\hbar v_{\text{eff}}^0 |\boldsymbol{k^* }- \boldsymbol{K}|)$.

\begin{figure}
    \centering
    \includegraphics[width=\linewidth]{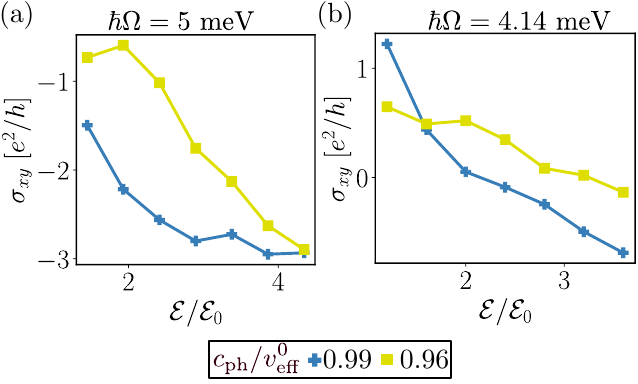}
    \caption{Comparing the dependence of $\sigma_{xy}$ on $\mathcal{E}$ for (a) the frequency considered in the main text and (b) a lower frequency where Floquet-Umklapp processes are stronger. Note that the frequency in panel (b) is inaccessible without generating two-photon resonances in the continuum model due to the peaked shape of the $\nu = \pm 1$ bands near the $\Gamma$ point.}
    \label{fig:fig_compare_freq}
\end{figure}

\section{Different Frequencies} \label{sec:diff-frequencies}
Reducing $\Omega$ below the value considered above will increase the ratio $(v_F e \mathcal{E}/\Omega^2)^2$ and in turn strengthen Floquet Umklapp processes, modifying the shape of the $\sigma_{xy}$ curve. We demonstrate this in Fig. \ref{fig:fig_compare_freq}(b) for an angular frequency $\Omega = 4.135 \ \mathrm{meV}/\hbar$. However, such a low-frequency regime is inaccessible in the continuum model (without generating two-photon resonances) due to the peaked shape of the continuum model $\nu = \pm 1$ band near the $\Gamma$ point, so we do not consider this lower (doubly-resonant) frequency regime in the main text.

\begin{figure}
    \centering
    \includegraphics[width=0.83\linewidth]{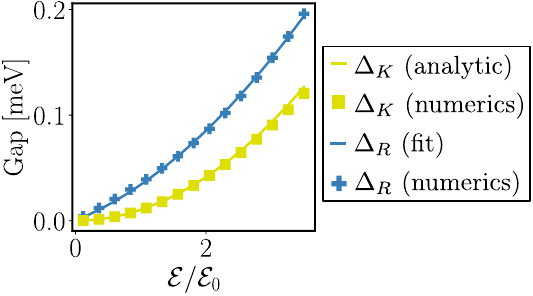}
    \caption{Comparison of the fitted $\Delta_R$ and predicted $\Delta_K$ in Equations \ref{eq:deltar} and \ref{eq:deltak} (solid lines) to those obtained from numerics (points), using $\hbar\Omega = 5 \ \mathrm{meV}$ in the toy model. Here, we fit $\Delta_R$ with factors of $f^{R}_{1} = 0.04$ and $f^{R}_{2} = 0.0184$ (see Eq. \ref{eq:deltar}).}
    \label{fig:fig_gaps}
\end{figure}

\section{Gap Sizes} \label{sec:gap-sizes}
In this section, we estimate the size of the Floquet-induced gaps $\Delta_K$ and $\Delta_R$. By the rotating wave approximation, the Floquet-induced gap at the resonance ring, $\Delta_R$, is roughly proportional to the drive energy \cite{Rudner2020BandInsulators}. For a resonant drive that couples electronic states near the Dirac points, the drive energy is roughly
\begin{equation} \label{eq:dr}
    v_F e A / \hbar,
\end{equation}
as predicted by minimal coupling $\boldsymbol{q} \to \boldsymbol{q} + e \boldsymbol{A}(t) / \hbar$ in the Dirac cone Hamiltonian
\begin{equation}
    H_K(\boldsymbol{q}) = \hbar v_F \boldsymbol{q} \cdot (\xi \sigma_x, \sigma_y)
\end{equation}
with $v_F = W L_M / (2\sqrt{3} \hbar)$. 
(We always use perturbative drives that generally fall in the range of $\tilde{\mathcal{E}} < 1$.) We expect that 
\begin{equation}
    \Delta_R \approx \frac{\hbar v_F}{L_M} \tilde{\mathcal{E}}.
\end{equation}
Such an approximation works well for low-frequency resonant drives that couple states near the Dirac points. However, resonant drives with higher frequencies, like those used in the main text, couple states closer to the $\Gamma$-points of the TBG energy dispersion where the bands are nonlinear in $q$. In such a case, higher order (e.g., O($\tilde{\mathcal{E}}^2$)) contributions (from $O(q^2)$ contributions of the bandstructure) to $\Delta_R$ become dominant. In the present example, the energy of the tight binding model for graphene is quadratic in momentum near the $\Gamma$ point, so we write an ansatz
\begin{equation} \label{eq:deltar}
    \Delta_R \approx \frac{\hbar v_F}{L_M} (f^{R}_{1} \tilde{\mathcal{E}} + f^{R}_{2} \tilde{\mathcal{E}}^2),
\end{equation}
and fit $f^{R}_{1}$ and $f^{R}_{2}$ to match $\Delta_R$ obtained by numerically diagonalizing the Floquet Hamiltonian, as shown in Fig. \ref{fig:fig_gaps}. 

We can estimate the Floquet-induced $K$-point gap, $\Delta_K$, by considering the time-dependent Dirac Hamiltonian
\begin{align}
\begin{split}
    H_K(\boldsymbol{q},t) &= \hbar v_F (\xi  q_x \sigma^x + q_y \sigma^y) \\
    &+ v_F e A [ \xi \cos(\Omega t) \sigma^x -\sin(\Omega t) \sigma^y].
\end{split}
\end{align}
and performing a Van Vleck expansion \cite{feete-review,kitagawa,Rudner2020BandInsulators} to obtain an effective Floquet Hamiltonian
\begin{equation} \label{eq:van-vleck}
    H_{K,\text{eff}}(\boldsymbol{q}) = H_K^{(0)} + \frac{[H_K^{(-1)}, H_K^{(1)}]}{\hbar \Omega} = H_K + \xi \frac{e^2 v_F^2 A^2}{\hbar\Omega} \sigma^z
\end{equation}
with
\begin{equation}
    H^{(n)}_K(\boldsymbol{q}) = \frac{1}{2\pi /\Omega} \int_0^{2\pi/\Omega} H_K(\boldsymbol{q},t) e^{-in\Omega t} dt.
\end{equation}
From Eq. \ref{eq:van-vleck}, we can extract
\begin{equation} \label{eq:deltak}
    \Delta_K = \frac{2 e^2 v_F^2}{\hbar \Omega} A^2 = \frac{6\hbar v_F^2}{L_M^2 \Omega} \tilde{\mathcal{E}}^2.
\end{equation}

\section{Floquet Boltzmann Equation} \label{sec:sup-fbe}
Here, we present the full expression for the Floquet-Boltzmann equation \cite{fbe_adv}, $\partial_t F_{\boldsymbol{k}\alpha}(t) = I^{\text{el-ph}}_{\boldsymbol{k}\alpha}[\{ F_{\boldsymbol{k}\alpha}(t) \}] + I^{\text{el-el}}_{\boldsymbol{k}\alpha}[\{ F_{\boldsymbol{k}\alpha}(t) \}]$. The electron-phonon collision integral is
\begin{widetext}
\begin{align} \label{eq:el-ph-i}
\begin{split}
    I^{\text{el-ph}}_{\boldsymbol{k}\alpha}&[\{ F_{\boldsymbol{k}\alpha} \}] =  \frac{2\pi}{\hbar} \frac{1}{N} \sum_{{\boldsymbol{k}'\in \mathrm{BZ}}} \sum_{\alpha'}  \sum_j \sum_n |\mathcal{G}_{\boldsymbol{k}\alpha}^{\boldsymbol{k}'\alpha'}(n,j)|^2  \\
    & \hspace{1.7cm} \times \Big[ \left\{ F_{\boldsymbol{k}'\alpha'}(1- F_{\boldsymbol{k}\alpha}) \mathcal{N}(\hbar\omega_j(\boldsymbol{k}'-\boldsymbol{k})) - F_{\boldsymbol{k}\alpha} (1-F_{\boldsymbol{k}'\alpha'}) [1 + \mathcal{N}(\hbar\omega_j(\boldsymbol{k}'-\boldsymbol{k}))] \right\} \\
    & \hspace{3cm} \times \delta(\varepsilon_{\boldsymbol{k}'\alpha'} - \varepsilon_{\boldsymbol{k}\alpha} + \hbar \omega_j(\boldsymbol{q}) + n\hbar\Omega) \\
    & \hspace{2cm} + \left\{ F_{\boldsymbol{k}'\alpha'}(1- F_{\boldsymbol{k}\alpha}) [1 + \mathcal{N}(\hbar\omega_j(\boldsymbol{k}'-\boldsymbol{k}))] - F_{\boldsymbol{k}\alpha} (1-F_{\boldsymbol{k}'\alpha'}) \mathcal{N}(\hbar\omega_j(\boldsymbol{k}'-\boldsymbol{k})) \right\} \\
    & \hspace{3cm} \times \delta(\varepsilon_{\boldsymbol{k}'\alpha'} - \varepsilon_{\boldsymbol{k}\alpha} -\hbar \omega_j(\boldsymbol{q}) + n\hbar\Omega) \Big]
\end{split}
\end{align}
\begin{equation} \label{eq:el-ph-m}
    \Ch{\mathcal{G}_{\boldsymbol{k}\alpha}^{\boldsymbol{k}'\alpha'}(n,j) =  \frac{1}{\sqrt{A_{\text{Moir{\'e}}}}} \frac{D}{\sqrt{2\rho} c_{\text{ph}}}  \sqrt{\hbar \omega_j(\boldsymbol{k}'-\boldsymbol{k} )} \sum_m \sum_{\nu,\nu'} \langle \phi^{n+m}_{\boldsymbol{k}'\alpha'} | \nu'\boldsymbol{k}' \rangle \mathcal{W}^{\xi\nu'\nu}_{\boldsymbol{k},\boldsymbol{k}+\boldsymbol{G}_j}\langle \nu\boldsymbol{k} | \phi^m_{\boldsymbol{k}\alpha} \rangle}
\end{equation}
\end{widetext}
where $\rho = \Ch{7.61} \times 10^{-7} \ \mathrm{kg/m^2}$ is the 2D density of the graphene layers, $D$ is the deformation potential, and the acoustic phonon mode $j$ has frequency $\omega_j(\boldsymbol{q}) = \hbar c_{\text{ph}} |\boldsymbol{q} + \boldsymbol{G}_j|$ with $\{\boldsymbol{G}_j\}$ being the set of all possible reciprocal lattice vectors. The function $\mathcal{N}(\varepsilon) = (e^{-\varepsilon/k_B T_{\text{ph}}} - 1)^{-1}$ is the Bose-Einstein occupation of the phonon bath at temperature $T_{\text{ph}}$. The electron-electron collision integral is 
\begin{widetext}
\begin{align} \label{eq:el-el-i}
\begin{split}
    I^{\text{el-el}}_{\boldsymbol{k}\alpha}[\{ F_{\boldsymbol{k}\alpha} \}] = \frac{4\pi}{\hbar} &\frac{1}{N^2} \sum_{\boldsymbol{k}_2 \in \mathrm{BZ}} \sum_{\boldsymbol{k}_3 \in \mathrm{BZ}} \sum_{\alpha_2, \alpha_3, \alpha_4} \sum_n \sum_{\boldsymbol{G}} | \mathcal{V}_{(\boldsymbol{k},\alpha), (\boldsymbol{k}_2,\alpha_2)}^{(\boldsymbol{k}_3,\alpha_3),(\boldsymbol{k}_1+\boldsymbol{k}_2-\boldsymbol{k}_3,\alpha_4)}(n,\boldsymbol{G}) |^2 \times \\
    &\times \delta(\varepsilon_{\boldsymbol{k}\alpha} +  \varepsilon_{\boldsymbol{k}_2\alpha_2} - \varepsilon_{\boldsymbol{k}_3\alpha_3} - \varepsilon_{\boldsymbol{k} + \boldsymbol{k}_2-\boldsymbol{k}_3,\alpha_4} + n\hbar \Omega)  \times\\
    & \times\left[ (1-F_{\boldsymbol{k}\alpha})(1-F_{\boldsymbol{k}_2\alpha_2}) F_{\boldsymbol{k}_3 \alpha_3} F_{\boldsymbol{k}_1+\boldsymbol{k}_2-\boldsymbol{k}_3,\alpha_4} - F_{\boldsymbol{k}\alpha}F_{\boldsymbol{k}_2\alpha_2} (1-F_{\boldsymbol{k}_3 \alpha_3}) (1-F_{\boldsymbol{k}_1+\boldsymbol{k}_2-\boldsymbol{k}_3,\alpha_4} )\right]
\end{split}
\end{align}
\begin{align}
\begin{split}
    \mathcal{V}_{(\boldsymbol{k},\alpha), (\boldsymbol{k}_2,\alpha_2)}^{(\boldsymbol{k}_3,\alpha_3),(\boldsymbol{k}_1+\boldsymbol{k}_2-\boldsymbol{k}_3,\alpha_4)}(n) = \sum_{\nu_1,\nu_2} \sum_{\nu_3,\nu_4} \sum_{n_2,n_3,n_4} & V_{\boldsymbol{k}_2 -\boldsymbol{k}_3+\boldsymbol{G}} \Ch{\mathcal{W}^{\xi\nu_1\nu_4}_{\boldsymbol{k}_1,\boldsymbol{q}+\boldsymbol{G}} \mathcal{W}^{\xi\nu_2\nu_3}_{\boldsymbol{k}_2,-\boldsymbol{q}-\boldsymbol{G}}} \langle \phi^{n-n_2 + n_3 + n_4}_{\boldsymbol{k} \alpha} | \nu_1 \boldsymbol{k} \rangle  \langle \phi^{n_2}_{\boldsymbol{k}_2 \alpha_2} | \nu_2 \boldsymbol{k}_2 \rangle \times \\ 
    & \times \langle \nu_3 \boldsymbol{k}_3 | \phi^{n_3}_{\boldsymbol{k}_3 \alpha_3} \rangle \langle \nu_4 \boldsymbol{k}_4 | \phi^{n_4}_{\boldsymbol{k} + \boldsymbol{k}_2 - \boldsymbol{k}_3, \alpha_4} \rangle.
\end{split}
\end{align}
\end{widetext}
We solve for $\partial_t F_{\boldsymbol{k}\alpha} = 0$ using the Newton-Raphson algorithm. To ensure charge neutrality, we add a Lagrange multiplier term $\lambda (\sum_{\boldsymbol{k}\alpha} F_{\boldsymbol{k}\alpha} - N)$ to the Floquet-Boltzmann equation, choosing some large constant $\lambda$.

{

\color{black}
\section{Form Factor Details} \label{sec:form-factor}
Here, we discuss the details of the form factors $\mathcal{W}^{\xi\nu'\nu}_{\boldsymbol{k},\boldsymbol{q}+\boldsymbol{G}}$ used in the Boltzmann equation. In the continuum model, we calculate the form factor directly from the wavefunctions:
\begin{equation}
    \mathcal{W}^{\xi\nu'\nu}_{\boldsymbol{k},\boldsymbol{q}+\boldsymbol{G}} = \sum_{X,\boldsymbol{G}'} C^{X*}_{\nu'\boldsymbol{k}+\boldsymbol{q}} (\boldsymbol{G}'-\boldsymbol{G}) C^{X}_{\nu\boldsymbol{k}}(\boldsymbol{G}').
\end{equation}
For the toy model, we include, by hand, a suppression factor $e^{-l_w^2 |\boldsymbol{q}+\boldsymbol{G}|^2/4}$, which accounts for the moir{\'e} periodicity that the toy model is unable to capture \cite{linphonon}:
\begin{equation}
    \mathcal{W}^{\xi\nu'\nu}_{\boldsymbol{k},\boldsymbol{q}+\boldsymbol{G}} = \langle \xi \nu' \boldsymbol{k} + \boldsymbol{q} |\xi \nu \boldsymbol{k} \rangle e^{-l_w^2 |\boldsymbol{q}+\boldsymbol{G}|^2/4}.
\end{equation}
We choose $l_w = L_M / (1.5 \sqrt{3})$ so that the form factor dependence on $|\boldsymbol{q}+\boldsymbol{G}|$ captures that of the continuum model. 
\begin{figure*}
    \centering
    \includegraphics[width=0.7\linewidth]{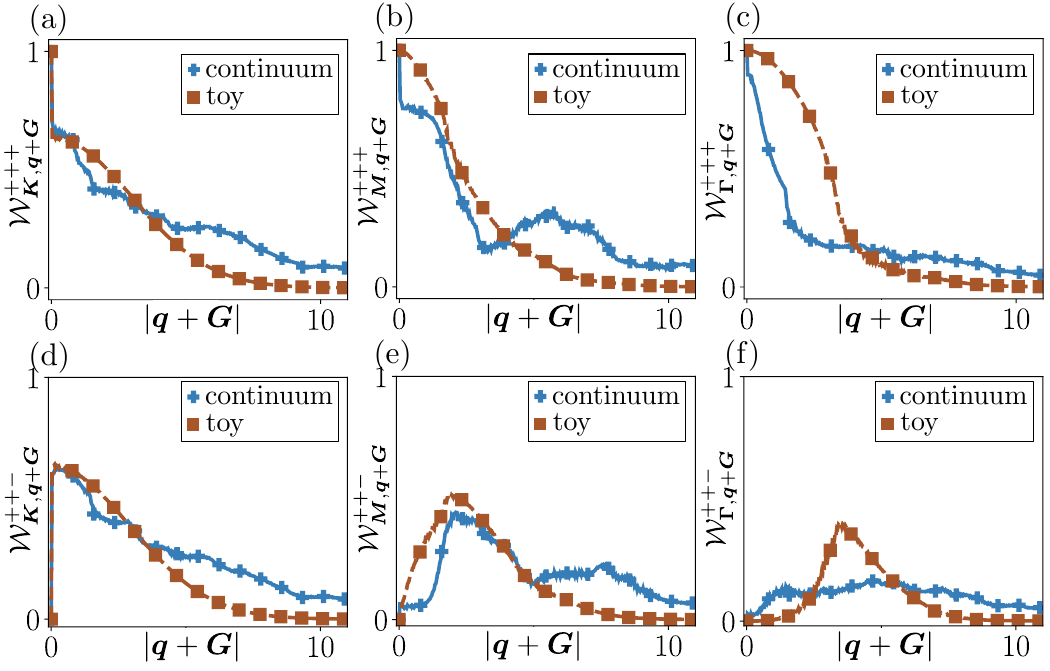}
    \caption{\Ch{Comparing the average toy and continuum model form factors $\mathcal{W}^{\xi\nu'\nu}_{\boldsymbol{k},\boldsymbol{q}+\boldsymbol{G}}$, with $|\boldsymbol{q} + \boldsymbol{G}|$ measured in units of $10^8 \ \mathrm{m^{-1}}$. (a-c) The intraband ($\nu = \nu'$) form factors at the $K$, $M$, and $\Gamma$ points. (d-f) The interband ($\nu \neq \nu'$) form factors at the $K$, $M$, and $\Gamma$ points.}}
    \label{fig:fig_formfactor}
\end{figure*}

To check the agreement between the toy and continuum model form factors, we calculate the form factors on a $\boldsymbol{k}$-grid, with a set of reciprocal lattice vectors $\{\boldsymbol{G}\}$, and plot the average value of $\mathcal{W}^{\xi\nu'\nu}_{\boldsymbol{k},\boldsymbol{q}+\boldsymbol{G}}$ as a function of $|\boldsymbol{q}+\boldsymbol{G}|$, for $\boldsymbol{k}$ at the $K$, $\Gamma$, and $M$ points of the mini Brillouin zone. The results, shown in Fig. \ref{fig:fig_formfactor}, show that the toy model form factor captures the general features of that of the continuum model. 

We can show analytically that the toy and continuum model form factors agree when $\boldsymbol{k}$ and $\boldsymbol{k}+\boldsymbol{q}$ lie within the same $K$ or $K'$ valley in the mini Brillouin zone. These low-momentum transfer processes are crucial to the tunable $\sigma_{xy}$ presented in the text. Let us write $\boldsymbol{k} \equiv \boldsymbol{K} + \boldsymbol{p}$ (and $\boldsymbol{k}+\boldsymbol{q}= \boldsymbol{K} + \boldsymbol{p} + \boldsymbol{q}$), where $|\boldsymbol{q}|$ and $|\boldsymbol{p}|$ are small enough such that the energy dispersion still resembles a Dirac cone at momenta $\boldsymbol{k}$ and $\boldsymbol{k}+\boldsymbol{q}$. By utilizing the eigenfunctions of the Dirac cone Hamiltonian, one can show that the form factors are
\begin{equation}
    \mathcal{W}_{\boldsymbol{K}+\boldsymbol{p},\boldsymbol{q}}^{\xi\nu'\nu} = \frac{1}{2} \left(1 + \nu \nu' \frac{\boldsymbol{p} \cdot (\boldsymbol{p} + \boldsymbol{q})}{|\boldsymbol{p}||\boldsymbol{p} +  \boldsymbol{q}|} \right).
\end{equation}
The formula holds true regardless of the chirality of the Dirac nodes. Note that form factors representing direct (non-FU) scattering transitions between different $K$ and $K'$ valleys in the mini Brillouin zone are not relevant since such scattering transitions are kinematically prohibited due to the slow-electron regime.

\section{Validity of the Diagonal Density Matrix Approximation}
In general, one needs to keep track of all coherences between the Floquet states, $\langle \hat{f}^{(\xi)\dagger}_{\boldsymbol{k}\alpha} (t) \hat{f}_{\boldsymbol{k}'\alpha'}^{(\xi)} (t) \rangle$, to fully-characterize the steady-state of a Floquet system. Translation symmetry suppresses the coherences for $\boldsymbol{k}\neq \boldsymbol{k}'$. The $\alpha \neq \alpha'$ interband coherences are suppressed for  $\tau_{\boldsymbol{k}}^{\text{tot}} \gg \hbar / \Delta \varepsilon_{\boldsymbol{k}}$, where $1/\tau_{\boldsymbol{k}}^{\text{tot}} = 1/\tau_{\boldsymbol{k}}^{\text{el}}+ 1/\tau_{\boldsymbol{k}}^{\text{ph}}$,  $1/\tau_{\boldsymbol{k}}^{\text{el}}$ and $ 1/\tau_{\boldsymbol{k}}^{\text{ph}}$ are the interband electron-electron and electron-phonon scattering rates, respectively, and $\Delta \varepsilon_{\boldsymbol{k}} = \min_{n\in \mathbb{Z}} |\varepsilon_{\boldsymbol{k}+} - \varepsilon_{\boldsymbol{k}-} + n\hbar\Omega|$. In this section, we will explain how we numerically estimate the scattering rates.

\subsection{Formal Definition of Scattering Times}
Following Ref. \cite{fbe_orig}, we define the interband scattering rates (fixing the initial Floquet band $\alpha$) as 
\begin{widetext}
\begin{align} \label{eq:el-ph-scat-time}
\begin{split}
    \frac{1}{\tau^{\text{ph}}_{\boldsymbol{k}}}&=  \frac{2\pi}{\hbar} \frac{1}{N} \sum_{{\boldsymbol{k}'\in \mathrm{BZ}}} \sum_{\alpha' \neq \alpha}  \sum_j \sum_n |\mathcal{G}_{\boldsymbol{k}\alpha}^{\boldsymbol{k}'\alpha'}(n,j)|^2  \Big[  (1-F_{\boldsymbol{k}'\alpha'}) [1 + \mathcal{N}(\hbar\omega_j(\boldsymbol{k}'-\boldsymbol{k}))] \delta(\varepsilon_{\boldsymbol{k}'\alpha'} - \varepsilon_{\boldsymbol{k}\alpha} + \hbar \omega_j(\boldsymbol{q}) + n\hbar\Omega) \\
    & + (1-F_{\boldsymbol{k}'\alpha'}) \mathcal{N}(\hbar\omega_j(\boldsymbol{k}'-\boldsymbol{k}))  \delta(\varepsilon_{\boldsymbol{k}'\alpha'} - \varepsilon_{\boldsymbol{k}\alpha} -\hbar \omega_j(\boldsymbol{q}) + n\hbar\Omega) \Big],
\end{split}
\end{align}
\begin{align} \label{eq:el-el-scat-time}
\begin{split}
     \frac{1}{\tau^{\text{el}}_{\boldsymbol{k}}} = \frac{4\pi}{\hbar} &\frac{1}{N^2} \sum_{\boldsymbol{k}_2 \in \mathrm{BZ}} \sum_{\boldsymbol{k}_3 \in \mathrm{BZ}} \sum_{\substack{\alpha_2, \alpha_3, \alpha_4 \\ \alpha_3 \text{ or } \alpha_4 \neq \alpha}} \sum_n \sum_{\boldsymbol{G}} | \mathcal{V}_{(\boldsymbol{k},\alpha), (\boldsymbol{k}_2,\alpha_2)}^{(\boldsymbol{k}_3,\alpha_3),(\boldsymbol{k}_1+\boldsymbol{k}_2-\boldsymbol{k}_3,\alpha_4)}(n,\boldsymbol{G}) |^2 \times \\
    &\times \delta(\varepsilon_{\boldsymbol{k}\alpha} +  \varepsilon_{\boldsymbol{k}_2\alpha_2} - \varepsilon_{\boldsymbol{k}_3\alpha_3} - \varepsilon_{\boldsymbol{k} + \boldsymbol{k}_2-\boldsymbol{k}_3,\alpha_4} + n\hbar \Omega) F_{\boldsymbol{k}_2\alpha_2} (1-F_{\boldsymbol{k}_3 \alpha_3}) (1-F_{\boldsymbol{k}_1+\boldsymbol{k}_2-\boldsymbol{k}_3,\alpha_4} ).
\end{split}
\end{align}
\end{widetext}

\subsection{Numerical Calculation of Scattering Rates} \label{sec:numerical-scat-rate}
In Fig. \ref{fig:fig_scattime}, we show the ratio $\hbar / ( \tau^{\text{tot}}_{\boldsymbol{k}}\Delta \varepsilon_{\boldsymbol{k}})$ in the regimes $\mathcal{E}/\mathcal{E}_0  = 0.966 < \mathcal{E}^*/\mathcal{E}_0 $ and $\mathcal{E}/\mathcal{E}_0  = 2.898  > \mathcal{E}^*/\mathcal{E}_0 $ for the case $\chi = \tau^{\text{el}}_K / \tau^{\text{ph}}_K \approx 2.8$ and $\zeta = \hbar / (2\tau^{\text{tot}}_K \Delta_K) \approx 0.5$, where $\zeta$ is the maximum value across the range of $\mathcal{E}$ considered in Fig. 1(b) in the main text, and $\chi$ is evaluated at the drive amplitude at which $\zeta$ is fixed. One sees that $\hbar / ( \tau^{\text{tot}}_{\boldsymbol{k}}\Delta \varepsilon_{\boldsymbol{k}}) \ll 1$ for most of the Brillouin zone and $\hbar / ( \tau^{\text{tot}}_{\boldsymbol{k}}\Delta \varepsilon_{\boldsymbol{k}}) < 1$ where the interband gaps are the smallest; by analysis of the Floquet-Redfield equation in Refs. \cite{fbe_orig,k_thesis}, the diagonal density matrix was shown to be a good approximation in this regime. Note that the definitions in Eqs. \ref{eq:el-ph-scat-time} and \ref{eq:el-el-scat-time} calculate the electron (rather than hole) scattering times, and hence Pauli blocking results in different scattering times for the upper and lower Floquet bands; the scattering rates quoted in the main text take the maximum rate. Separately, we also note that the toy model underestimates the resonance gap (see Fig. \ref{fig:fig_bands}) relative to the continuum model, and therefore overestimates $\hbar / ( \tau^{\text{tot}}_{\boldsymbol{k}}\Delta \varepsilon_{\boldsymbol{k}})$ around the resonance ring.

\begin{figure}[b]
    \centering
    \includegraphics[width=\linewidth]{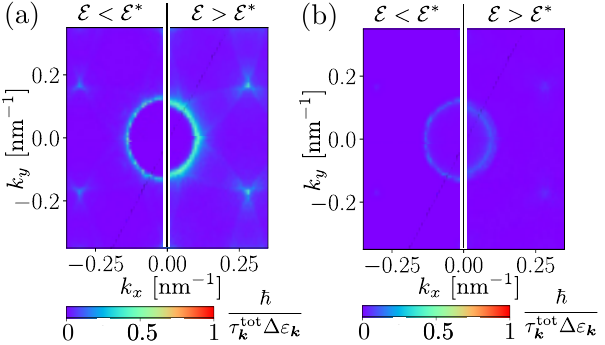}
    \caption{\Ch{The color represents the ratio $ \hbar/(\tau^{\text{tot}}_{\boldsymbol{k}}\Delta \varepsilon_{\boldsymbol{k}})$ at different points in momentum space (a) for the upper Floquet band and (b) for the lower Floquet band for the case $\chi \approx 2.8$ and $\zeta \approx 0.5$.}}
    \label{fig:fig_scattime}
\end{figure}

\section{Dielectric Function of Strontium Titanate}
The dielectric function for $\mathrm{SrTiO_3}$ is
\begin{equation}
    \epsilon(\Omega) = \epsilon_{\infty} \prod_{j = 1}^{3} \frac{\omega_{Lj}^2 - \Omega^2}{\omega_{Tj}^2 - \Omega^2}
\end{equation}
at angular frequency $\Omega$, with the experimentally-determined longitudinal and transverse optical phonon frequencies, $\omega_{Lj}$ and $\omega_{Tj}$, respectively, given in Ref. \cite{PhysRevB.94.224515, PhysRevB.24.3086,doi:10.1063/1.363513}. One finds that $|\epsilon(5 \ \mathrm{meV})| = 1682$. 
}

\section{Monkhorst-Pack Grid, Numerical Integration, and Convergence} \label{sec:monkhorst-pack-sec}
In this section, we describe the methods we use to discretize the momentum Brillouin zone. We perform the Boltzmann equation integrals, introduced in Equations \ref{eq:el-ph-i} and \ref{eq:el-el-i}, over an $N \times N$ Monkhorst-Pack (MP) set of grid points \cite{mp-grid}, with $\boldsymbol{k}$-points
\begin{equation}
    \boldsymbol{k}_{m,n} = \frac{m\boldsymbol{G}_1 + n\boldsymbol{G}_2}{N},
\end{equation}
odd $N$, and $m, n = 0, \hdots, N - 1$. Specifically, we avoid values of $N (\mathrm{mod} \ 3) =0$ that generate a $\boldsymbol{k}$-point exactly at the high-symmetry point of $K$, because such grids converge poorly when the drive strength is weak and Floquet-induced gap $\Delta_K$ is small.
\begin{figure}
    \centering
    \includegraphics[width=\linewidth]{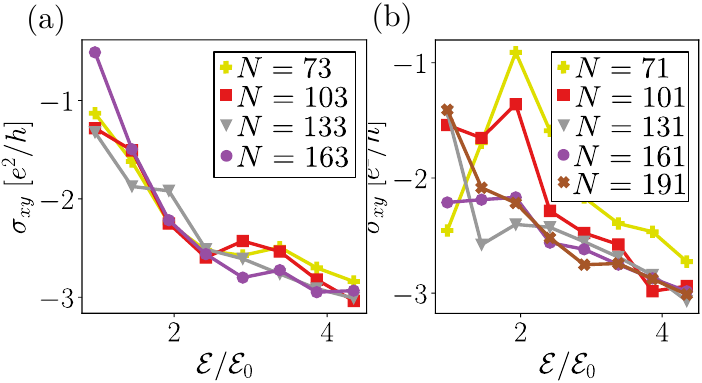}
    \caption{Convergence of anomalous conductivity with grid size for (a) $N (\text{mod } 3) = 1$ and (b) $N (\text{mod } 3) = 2$. Due to the positioning of grid points near the $K$ point, the results at low grid resolutions show significant disagreement.}
    \label{fig:fig_convergence}
\end{figure}

\subsection{Energy and Momentum Conservation} \label{sec:en-mom-conservation}
Here, we discuss in detail how we impose momentum and energy conservation on this MP grid. The space of MP $\boldsymbol{k}$ vectors are closed under addition and subtraction (modulo a reciprocal lattice vector), so conservation of momentum (e.g., $\boldsymbol{k} + \boldsymbol{k}_2 -\boldsymbol{k}_3$ in Eq. \ref{eq:el-el-i}), is simple to implement. We impose energy conservation via a smeared Dirac Delta function
\begin{equation}
    \delta(\varepsilon) = \begin{cases}
        1.04766 e^{-\varepsilon^2/2\sigma^2} / (2.5066283 \sigma), & \text{if } |\varepsilon| < 2\sigma, \\
        0, & \text{otherwise},
    \end{cases}
\end{equation}
where we have chosen numerical factors so that
\begin{equation}
    \int_{-\infty}^{\infty} \delta(\varepsilon) d\varepsilon = 1.
\end{equation}
The smearing parameter $\sigma$ is one-tenth of the maximum quasienergy spacing between nearest-neighbor MP $\boldsymbol{k}$-points
\begin{equation}
    \sigma = 0.1 \max_{\langle \boldsymbol{k},\boldsymbol{k}' \rangle, \alpha} |\varepsilon_{\boldsymbol{k}\alpha}^{(\xi)} - \varepsilon_{\boldsymbol{k}'\alpha}^{(\xi)}|,
\end{equation}
where $\langle \boldsymbol{k},\boldsymbol{k}' \rangle$ restricts $\boldsymbol{k}'$ to be a nearest-neighbor of $\boldsymbol{k}$, and we have tuned the prefactor of $0.1$ so that upon calculating the steady-state without Floquet-Umklapp processes, we obtain a Fermi-Dirac distribution, $F_{\boldsymbol{k}\alpha}^{(\xi)} = ({e^{\varepsilon_{\boldsymbol{k}\alpha}/k_BT_{\text{ph}}} + 1})^{-1}$
with temperature $T_{\text{ph}}$ of the phonon bath \cite{Galitskii1970ElectricWave}.  

\begin{figure}
    \centering
    \includegraphics[width=0.6\linewidth]{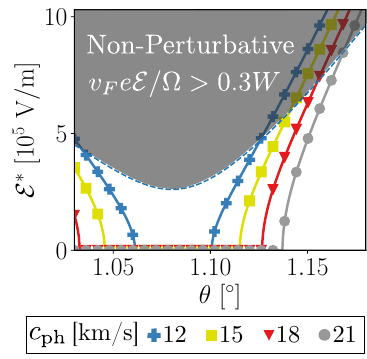}
    \caption{The requirement that the laser drive strength $\mathcal{E}$ is perturbative, i.e. a fraction of electron bandwidth $e\mathcal{E}L_M < W$, narrows the range of $\mathcal{E}$ values that can be used. As a result, the range of $c_{\text{ph}}$ whose $\mathcal{E}^*$ is visible is limited as well - we postulate that they are pushed to higher drive strengths $\mathcal{E}$.}
    \label{fig:limitations}
\end{figure}

\subsection{Convergence of Conductivities}
In Fig. \ref{fig:fig_convergence}, we show the convergence of the Hall conductivity $\sigma_{xy}$ with grid size, using $\hbar\Omega = 5 \ \mathrm{meV}$. In the main text, we use a $163 \times 163$ MP grid for non-interacting calculations, and a $73 \times 73$ grid for interacting calculations.

\section{Berry Curvature Calculations}
We follow the Berry curvature calculation presented in \cite{berry}, defining U(1) link variables
\begin{equation}
    U_{\mu}(\boldsymbol{k},t) = \frac{ \langle \alpha (\boldsymbol{k}, t) | \alpha(\boldsymbol{k} + \hat{\boldsymbol{\mu}}, t) \rangle }{|\langle \alpha (\boldsymbol{k}, t) | \alpha(\boldsymbol{k} + \hat{\boldsymbol{\mu}}, t) \rangle|}
\end{equation}
where $\mu = x,y$, $\hat{\boldsymbol{\mu}} = \boldsymbol{G}_{\mu} / N$, and $|\alpha(\boldsymbol{k},t) \rangle$ are the Bloch vectors (i.e., $| \psi_{\boldsymbol{k}\alpha} (t) \rangle = e^{-i\boldsymbol{k}\cdot \boldsymbol{r}} |\alpha (\boldsymbol{k},t) \rangle$). The Berry curvature is
\begin{equation}
    \mathcal{B}_{\boldsymbol{k}\alpha}(t) = \frac{(2\pi)^2}{N^2 A_{M} }\text{arg} \left[\frac{U_x(\boldsymbol{k},t) U_y(\boldsymbol{k}+\hat{\boldsymbol{x}},t)}{U_x(\boldsymbol{k}+\hat{\boldsymbol{y}},t) U_y(\boldsymbol{k},t)} \right]
\end{equation}
and we use the time-averaged Berry curvature
\begin{equation}
    \mathcal{B}_{\boldsymbol{k}\alpha} \equiv \frac{1}{2\pi/\Omega} \int_0^{2\pi/\Omega}  \mathcal{B}_{\boldsymbol{k}\alpha}(t) dt
\end{equation}
in transport calculations. 

\section{The Drive Amplitude Perturbative Regime at Different Twist Angles}
We have treated the laser drive as a perturbation to the undriven TBG Hamiltonian, which restricts the range of field strengths $\mathcal{E}$ we can use to a weak perturbative regime. This also narrows the range of phonon speeds $c_{\text{ph}}$ that will generate a critical field strength $\mathcal{E}^*$ in the perturbative regime, hence the narrow range of $c_{\text{ph}}$ we have considered in, e.g., Fig. 1(c). For various twist angles, we estimate the range of drive strengths $\mathcal{E}$ that are perturbative in the unshaded region of Fig. \ref{fig:limitations} and overlap in solid lines the predicted value of $\mathcal{E}^*$ for different speeds of sound. The shaded, non-perturbative regime corresponds to drive energy scales $v_F e \mathcal{E} / \Omega$ greater than a fraction, e.g., $0.3$, of the bandwidth $W$. Here, we follow the analysis in \cite{macdonald} to estimate the undriven Fermi velocity 
\begin{equation} \label{eq:v_f}
    v_F(\theta) = \sqrt{\left(({1-3\alpha^2})/({1+3\alpha^2 (1 + \eta^2)}) \times v_F^{\text{ml}}\right)^2 + v_{\text{min}}^2}, 
\end{equation}
where $v_{\text{min}} = 10^{4} \ \mathrm{m/s}$ is a manually set minimum Fermi velocity of the undriven flat bands, and we use the same parameters as in Sec. \ref{sec:choice}. We also adjust $\Omega$ such that $\Omega / v_F(\theta)$ is constant and equal to those considered in Figs. 1-4.

\bibliographystyle{apsrev4-1}